\documentclass[letterpaper,twocolumn,10pt]{article}
\usepackage{usenix_2020_09}

\usepackage{tikz}
\usepackage{amsmath}
\usepackage{booktabs}
\usepackage{adjustbox}
\usepackage{float}

\newcommand{\eg}{e.g.,~}
\newcommand{\ie}{i.e.,~}
\newcommand{\cf}{cf.~}
\newcommand{\etc}{etc.}
\newcommand{\etal}{et al.~}
\newcommand\mydots{\hbox to 1em{.\hss.\hss.}}

\begin{document}

\date{}

\title{\Large \bf Frontrunner Jones and the Raiders of the Dark Forest: \\ An Empirical Study of Frontrunning on the \\ Ethereum Blockchain}

\author{
{\rm Christof Ferreira Torres}\\
SnT, University of Luxembourg 
\and
{\rm Ramiro Camino}\\
Luxembourg Institute of Science and Technology
\and
{\rm Radu State}\\
SnT, University of Luxembourg
} %

\maketitle

\begin{abstract}
Ethereum prospered the inception of a plethora of smart contract applications, ranging from gambling games to decentralized finance.
However, Ethereum is also considered a highly adversarial environment, where vulnerable smart contracts will eventually be exploited.
Recently, Ethereum's pool of pending transaction has become a far more aggressive environment.
In the hope of making some profit, attackers continuously monitor the transaction pool and try to frontrun their victims' transactions by either displacing or suppressing them, or strategically inserting their transactions. 
This paper aims to shed some light into what is known as a \emph{dark forest} and uncover these predators' actions.
We present a methodology to efficiently measure the three types of frontrunning: \emph{displacement}, \emph{insertion}, and \emph{suppression}.
We perform a large-scale analysis on more than 11M blocks and identify almost 200K attacks with an accumulated profit of 18.41M USD for the attackers, providing evidence that frontrunning is both, lucrative and a prevalent issue.
\end{abstract}

\section{Introduction}

The concept of frontrunning is not new. 
In financial markets, brokers act as intermediaries between clients and the market, and thus brokers have an advantage in terms of insider knowledge about potential future buy/sell orders which can impact the market. 
In this context, frontrunning is executed by prioritizing a broker's trading actions before executing the client's orders such that the trader pockets a profit. 
Frontrunning is illegal in regulated financial markets. 
However, the recent revolution enabled by decentralized finance (DeFi), where smart contracts and miners replace intermediaries (brokers) is both, a blessing and a curse.  
Removing trusted intermediaries can streamline finance and substantially lower adjacent costs, but misaligned incentives for miners leads to generalized frontrunning, in which market participants behave exactly like unethical brokers used to in the ``old'' financial world. 
Unfortunately, this is already happening at a large scale.
Our paper is among the first comprehensive surveys on the extent and impact of this phenomenon. 
Already in 2017, the Bancor ICO \cite{web_front_run0} was susceptible to such an attack -- among other vulnerabilities -- but no real attack was observed in the wild.  
Some concrete frontrunning attacks on the Ethereum blockchain were brought to knowledge by two independently reported attacks and their mitigation approaches to the informed audience. 
In the first report \cite{web_front_run2}, the researchers tried to recover some liquidity tokens by calling a specific function in a smart contract.
Since this function was callable by everyone, the authors -- who also compared the pending transactions in the transaction pool to a \emph{dark forest} full of predators -- assumed that their function call could be observed and frontrunned by bots observing the submitted transactions in the transaction pool. 
Even though they tried to obfuscate their efforts, their approach failed in the end, and they became a victim of a frontrunning bot. 
A few months later, a second group of researchers \cite{web_front_run5} reported a successful recovery using lessons learned from the previously reported incident \cite{web_front_run2}. 
The success was due to them mining their transactions privately without broadcasting them to the rest of the network. 
The researchers used a new functionality provided by SparkPool called the Taichi Network \cite{web_front_run6}. 
In this way, the transactions were not available to frontrunning bots but relied entirely on having a reliable and honest mining pool. However, this approach enables centralization and requires users to entrust their transactions to SparkPool.
Similar to how honeypots gather intelligence by luring attackers to compromise apparently vulnerable hosts \cite{Cheswick92anevening}, a recent experiment \cite{web_front_run7} detailed the interactions with two bots and reported relevant assessment on their nature and origin. 
Surprisingly, the frontrunning bots do not rely on advanced software development techniques or complex instructions, and code examples on developing such bots are readily available \cite{web_front_run1, web_front_run4}. 
There are several ways to perform frontrunning attacks. The first survey defining a taxonomy of frontrunning attacks \cite{eskandari2019sok} identified three different variants on how these can be performed. To understand these approaches -- \emph{displacement}, \emph{insertion}, and \emph{suppression} -- a short refresh on gas and transaction fees in Ethereum is given. 
Transactions, submitted to the Ethereum network, send money and data to smart contract addresses or account addresses. 
Transactions are confirmed by miners who get paid via a fee that the sender of the transaction pays. 
This payment is also responsible for the speed/priority miners include a transaction in a mined block. 
Miners have an inherent incentive to include high paying transactions and prioritize them. As such, nodes observing the unconfirmed transactions can frontrun by just sending transactions with higher payoffs for miners \cite{daian2019flash}. 
The common feature of all three attack types is that by frontrunning a transaction, the initial transaction's expected outcome is changed. 
In the case of the first attack (displacement), the outcome of a victim's original transaction is irrelevant. The second attack type (insertion) manipulates the victim's transaction environment, thereby leading to an arbitrage opportunity for the attacker. Finally, the third attack (suppression) delays the execution of a victim's original transaction. Although previous papers \cite{daian2019flash,eskandari2019sok} have identified decentralized applications which are victims of frontrunning attacks, no scientific study has analyzed the occurrence of these three attacks in the wild on a large scale.  
The impact of this structural design failure of the Ethereum blockchain is far-reaching. 
Many decentralized exchanges, implementing token-based market places have passed the 1B USD volume \cite{dex_status} and are prone to the same frontrunning attack vectors because the Ethereum blockchain is used as a significant building block. 
Frontrunning is not going to disappear any time soon, and the future looks rather grim. 
We do not expect to have mitigation against frontrunning in the short-term. 
Miners do profit from the fees and thus will always prioritize high yield transactions. 
Moreover, the trust mechanism in Ethereum is built on the total observability of the confirmed/unconfirmed transactions and is thus by design prone to frontrunning.
Our paper sheds light into the long term history of frontrunning on the Ethereum blockchain and provides the first large scale data-driven investigation of this type of attack vector.
We investigate the real profits made by attackers, differentiated by the specific attack type and propose the first methodology to detect them efficiently. 
\newline
\noindent
\textbf{Contributions.} We summarize our contributions as follows:

\noindent
\begin{itemize}
	\item We propose a methodology that is efficient enough to detect \emph{displacement}, \emph{insertion}, and \emph{suppression} attacks on Ethereum's past transaction history.
	\item We run an extensive measurement study and analyze frontrunning attacks on Ethereum for the past five years.
	\item We identify a total of 199,725 attacks, 1,580 attacker accounts, 526 bots, and over 18.41M USD profit.
	\item We demonstrate that the identified attacker accounts and bots can be grouped to 137 unique attacker clusters.
	\item We discuss frontrunning implications and find that miners made a profit of 300K USD due to frontrunners.
\end{itemize}

\begin{figure*}
  \centering
  \includegraphics[width=1.0\textwidth]{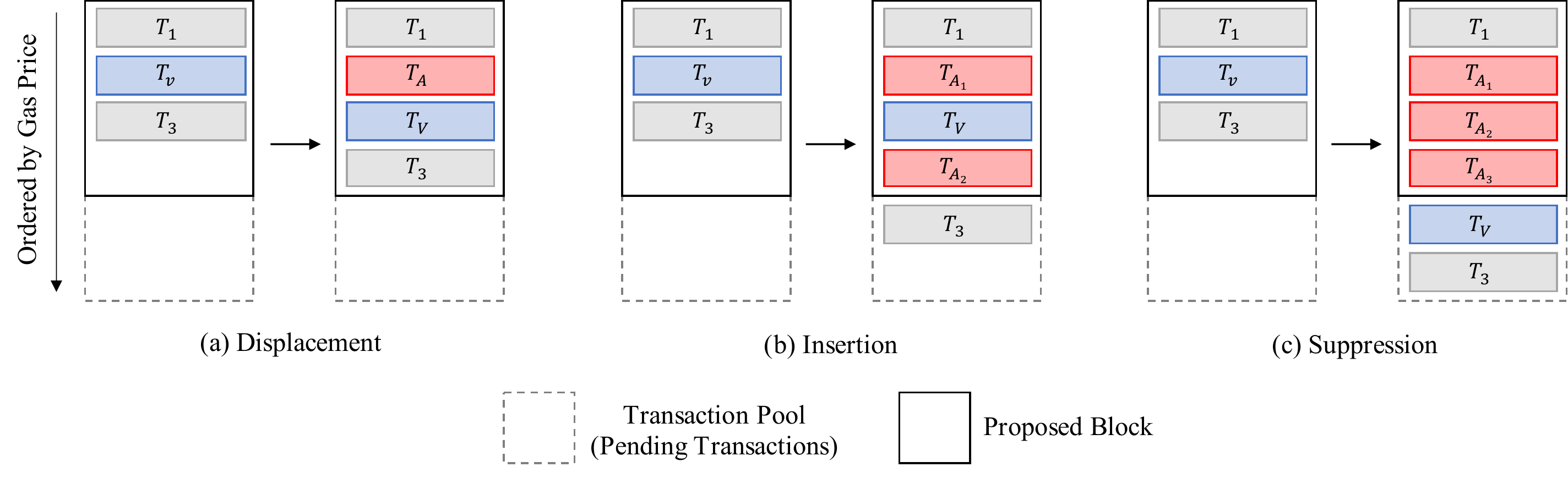}
  \caption{Illustrative examples of the three frontrunning attack types.}
  \label{fig:frontrunning}
\end{figure*}

\section{Background}

This section provides the necessary background to understand our work setting, including smart contracts, transactions, gas economics, and transaction ordering.

\subsection{Smart Contracts}

The notion of smart contracts has already been introduced in 1997 by Nick Szabo \cite{szabo1997}, but the concept only became a reality with the inception of Ethereum in 2015 \cite{wood2014ethereum}.
Ethereum proposes two types of accounts: externally owned accounts (EOA) and contract accounts (smart contracts).
EOAs are controlled via private keys and have no associated code. 
Contract accounts, \ie smart contracts, have associated code but are not controlled via private keys.
They operate as fully-fledged programs that are stored and executed across the blockchain.
EOAs and smart contracts are identifiable via a unique 160-bit address.
Smart contracts are immutable, and they cannot be removed or updated once they have been deployed unless they have been explicitly designed to do so.
Besides having a key-value store that enables them to preserve their state across executions, smart contracts also have a balance that keeps track of the amount of ether (Ethereum's cryptocurrency) that they own.
Smart contracts are usually developed using a high-level programming language, such as Solidity \cite{solidity}.
The program code is then compiled into a low-level bytecode representation, which is then interpreted by the Ethereum Virtual Machine (EVM). 
The EVM is a stack-based virtual machine that supports a set of Turing-complete instructions. 

\subsection{Transactions}

Smart contracts are deployed and executed via transactions.
Transactions contain an amount of ether, a sender, a receiver, input data, a gas limit and a gas price.
Transactions may only be initiated by EOAs.
Smart contract functions are invoked by encoding the function signature and arguments in a transaction's data field.
A fallback function is executed whenever the provided function name is not implemented.
Smart contracts may call other smart contracts during execution. Thus, a single transaction may trigger further transactions, so-called internal transactions. 

\subsection{Gas Economics}

Ethereum employs a gas mechanism that assigns a cost to each EVM instruction.
This mechanism prevents denial-of-service attacks and ensures termination.
When issuing a transaction, the sender has to specify a \emph{gas limit} and a \emph{gas price}. 
The gas limit is specified in gas units and must be large enough to cover the amount of gas consumed by the instructions during a contract's execution. 
Otherwise, the execution will terminate abnormally, and its effects will be rolled back.
The gas price defines the amount of ether that the sender is willing to pay per unit of gas used.
The sender is required to have a balance greater than or equal to gas limit $\times$ gas price, but
the final transaction fee is computed as the gas used $\times$ gas price.
The price of gas is extremely volatile as it is directly linked to the price of ether.
As a result, Breidenbach \etal \cite{gastoken} proposed GasToken, a smart contract that allows users to tokenize gas. 
The idea is to store gas when ether is cheap and spend it when ether is expensive, thereby allowing users to save on transaction fees.
Two versions of GasToken exist, whereby the second version is more efficient than the first one.
The first version of GasToken (GST1) exploits the fact that gas is refunded when storage is freed. 
Hence, gas is saved by writing to storage and liberated when deleting from storage. The second version of GasToken (GST2) exploits the refunding mechanism of removing contracts. Hence, gas is saved by creating contracts and liberated by deleting contracts.
In 2020, 1inch released their version of GST2 called ChiToken \cite{chigastoken}, which includes some optimizations.

\subsection{Transaction Ordering}

A blockchain is essentially a verifiable, append-only list of records in which all transactions are recorded in so-called blocks. This list is maintained by a distributed peer-to-peer (P2P) network of distrusting nodes called \emph{miners}. Miners follow a consensus protocol that dictates the appending of new blocks.
They compete to create a block by solving a cryptographic puzzle.
The winner is rewarded with a static block reward and the execution fees from the included transactions \cite{mining}.
While blockchains prescribe specific rules for consensus, there are only loose requirements for selecting and ordering transactions.
Thus, miners get to choose which transactions to include and how to order them inside a block.
Nevertheless, 95\% of the miners choose and order their transactions based on the gas price to increase their profit, thereby deliberately creating a prioritization mechanism for transactions \cite{zhou2020high}.

\section{Frontrunning Attacks}

This section defines our attacker model and introduces the reader to three different types of frontrunning attacks.

\subsection{Attacker Model}
\label{sec:attacker-model}

\begin{figure}
  \centering
  \includegraphics[width=1.0\linewidth]{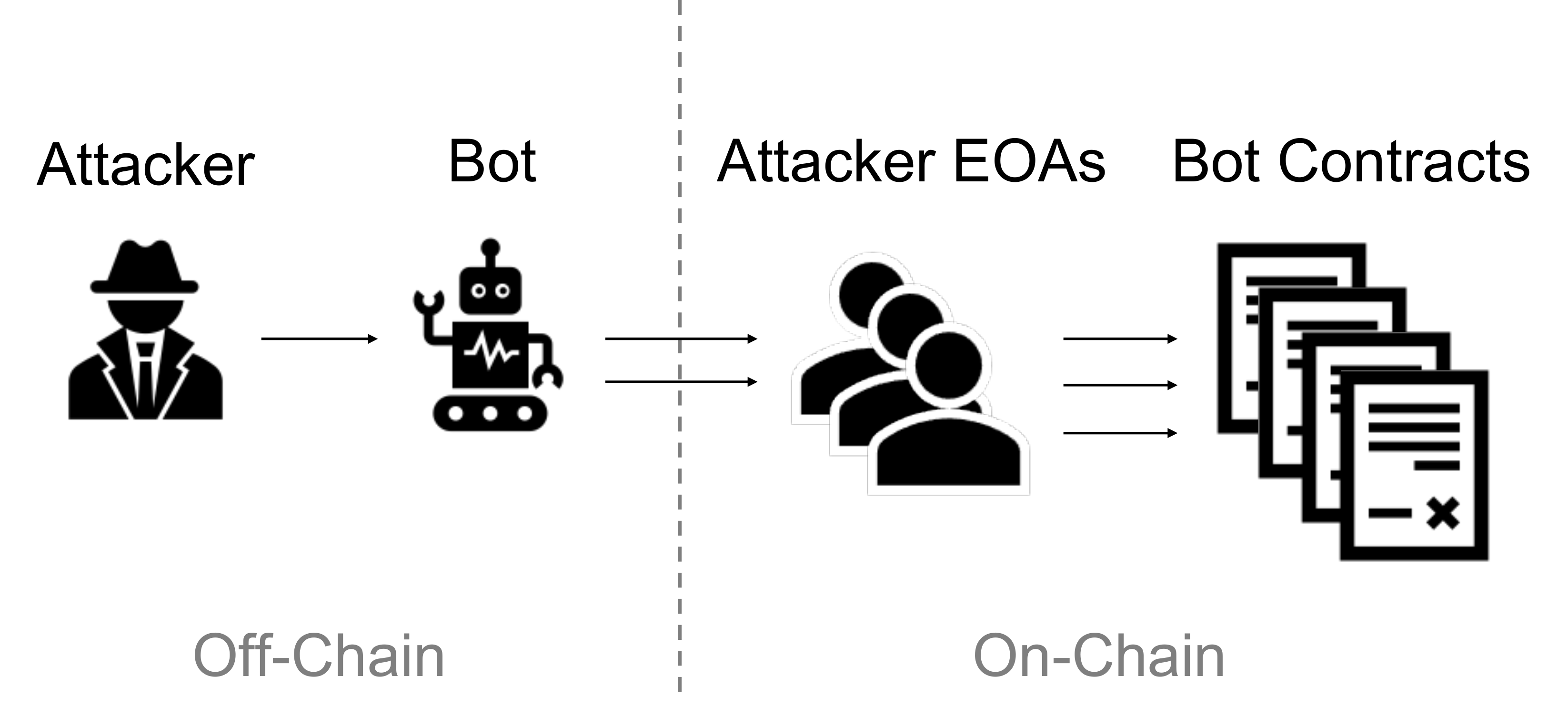}
  \caption{Attacker model with on-chain and off-chain parts.}
  \label{fig:attacker_model}
\end{figure}

Miners, as well as non-miners, can mount frontrunning attacks.
Miners are not required to pay a higher gas price to manipulate the order of transactions as they have full control over how transactions are included.
Non-miners, on the other hand, are required to pay a higher gas price in order to frontrun transactions of other non-miners.
Our attacker model assumes an attacker $A$ that is a financially rational non-miner with the capability to monitor the transaction pool for incoming transactions.
The attacker $A$ needs to process the transactions in the pool, find a victim $V$ among those transactions and create a given amount of attack transactions $T_{A_i}$ before the victim's transaction $T_V$ is mined.
Usually, $A$ would not be able to react fast enough to perform all these tasks manually.
Hence, we assume that the attacker $A$ has at least one computer program $Bot_A$ that automatically performs these tasks.
However, $Bot_A$ must be an off-chain program, because contracts cannot react on its own when transactions are added to the pool.
Nevertheless, $Bot_A$ needs at least one or more EOAs to act as senders of any attack transaction $T_A$.
Using multiple EOAs helps attackers obscure their frontrunning activities, similar to money laundering layering schemes.
We refer to these EOAs owned by $A$ as attacker accounts $EOA_{A_j}$ and to the EOA owned by $V$ as victim account $EOA_V$.
We assume that attacker $A$ owns a sufficiently large balance across all its attacker accounts $EOA_{A_j}$ from which it can send frontrunning transactions.
However, attacker $A$ can also employ smart contracts to hold part of the attack logic.
We refer to these smart contracts as bot contracts $BC_{A_k}$, which are called by the attacker accounts $EOA_{A_j}$.
Figure~\ref{fig:attacker_model} provides an overview of our final attacker model.

\subsection{Frontrunning Taxonomy}

We describe in the following the taxonomy of frontrunning attacks presented by Eskandari \etal \cite{eskandari2019sok}.

\begin{description}
\item [Displacement.] In a displacement attack an attacker $A$ observes a profitable transaction $T_V$ from a victim $V$ and decides to broadcast its own transaction $T_A$ to the network, where $T_A$ has a higher gas price than $T_V$ such that miners will include $T_A$ before $T_V$ (see Figure~\ref{fig:frontrunning} a).
Note that the attacker does not require the victim's transaction to execute successfully within a displacement attack. 
For example, imagine a smart contract that awards a user with a prize if they can guess the preimage of a hash. An attacker can wait for a user to find the solution and to submit it to the network. Once observed, the attacker then copies the user's solution and performs a displacement attack. The attacker's transaction will then be mined first, thereby winning the prize, and the user's transaction will be mined last, possibly failing.

\item [Insertion.]  In an insertion attack an attacker $A$ observes a profitable transaction $T_V$ from a victim $V$ and decides to broadcast its own two transactions $T_{A_1}$ and $T_{A_2}$ to the network, where $T_{A_1}$ has a higher gas price than $T_V$ and $T_{A_2}$ has a lower gas price than $T_V$, such that miners will include $T_{A_1}$ before $T_V$ and $T_{A_2}$ after $T_V$ (see Figure~\ref{fig:frontrunning} b). 
This type of attack is also sometimes called a \emph{sandwich attack}.
In this type of attack, the transaction $T_V$ must execute successfully as $T_{A_2}$ depends on the execution of $T_V$.
A well-known example of insertion attacks is arbitraging on decentralized exchanges, where an attacker observes a large trade, also known as a whale, sends a buy transaction before the trade, and a sell transaction after the trade.

\item [Suppression.] In a suppression attack, an attacker $A$ observes a transaction $T_V$ from a victim $V$ and decides to broadcast its transactions to the network, which have a higher gas price than $T_V$ such that miners will include $A$'s transaction before $T_V$ (see Figure~\ref{fig:frontrunning} c). The goal of $A$ is to suppress transaction $T_V$, by filling up the block with its transactions such that transaction $T_V$ cannot be included anymore in the next block. This type of attack is also called \emph{block stuffing}. Every block in Ethereum has a so-called \emph{block gas limit}. The consumed gas of all transactions included in a block cannot exceed this limit. 
$A$'s transactions try to consume as much gas as possible to reach this limit such that no other transactions can be included.
This type of attack is often used against lotteries where the last purchaser of a ticket wins if no one else purchases a ticket during a specific time window. Attackers can then purchase a ticket and mount a suppression attack for several blocks to prevent other users from purchasing a ticket themselves. Keep in mind that this type of frontrunning attack is expensive.

\end{description}

\section{Measuring Frontrunning Attacks}

This section provides an overview of our methodology's design and implementation details to detect frontrunning attacks in the wild.

\subsection{Identifying Attackers}
\label{sec:identifying-attackers}

As defined in Section~\ref{sec:attacker-model}, an attacker $A$ employs one or more off-chain programs to perform its attacks.
However, because we have no means to distinguish between the different software agents an attacker $A$ could have, for this study, we consider all of them as part of the same multi-agent system $Bot_A$.
Additionally, we cannot recognize the true nature of $A$ or how $Bot_A$ is implemented.
Instead, we would like to build a cluster with the $n$ different attacker accounts $EOA_{A_1}, \mydots, EOA_{A_n}$ and the $m$ different bot contracts $BC_{A_1}, \mydots, BC_{A_m}$ to form an identity of $A$.
Consequently, in each of the following experiments, we use our detection system's results to build a graph.
Each node is either an attacker account or a bot contract.
We make the following two assumptions:

\begin{description}
    \item[Assumption 1:] Attackers only use their own bot contracts.
    Hence, when an attacker account sends a transaction to a bot contract, we suspect that both entities belong to the same attacker.
    Note that one attacker account can send transactions to multiple bot contracts, and bot contracts can receive transactions from multiple attacker accounts.
    \item[Assumption 2:] Attackers develop their own bot contracts, and they do not publish the source code of their bot contracts as they do not want to share their secrets with competitors.
    Hence, when the bytecode of two bot contracts is exactly the same, we suspect that they belong to the same attacker.
\end{description}

\noindent
With these assumptions in mind, we create edges between attacker accounts and bot contracts that share at least one attack transaction, and between bots that share the same bytecode.
Using the resulting graph, we compute all the connected components.
Hence, we interpret each of these connected components as a single attacker cluster. 

\subsection{Detecting Displacement}

\begin{figure}
  \centering
  \includegraphics[width=0.4\textwidth]{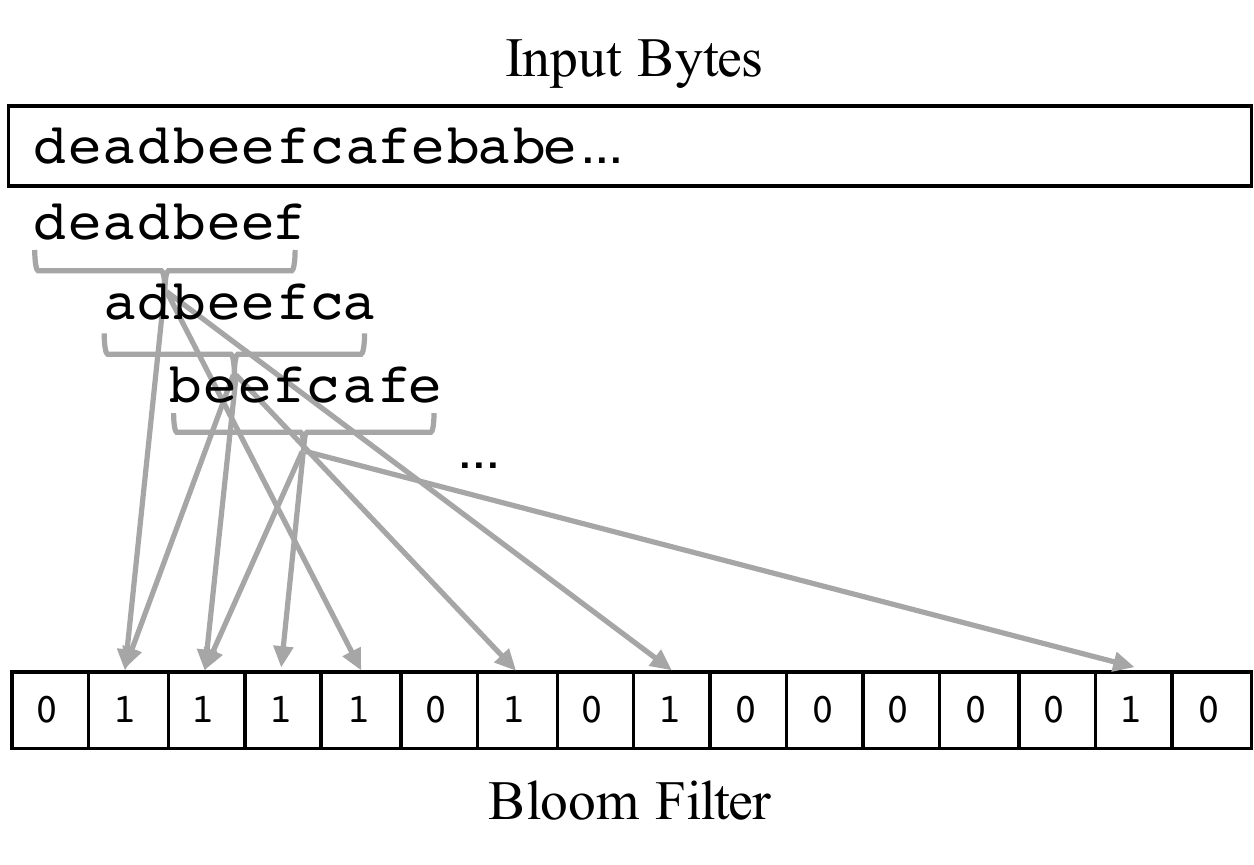}
  \caption{An example on how transaction input bytes are mapped into a bloom filter.}
  \label{fig:displacement_detection}
\end{figure}

Attackers typically perform displacement attacks by observing profitable pending transactions via the transaction pool and by copying these profitable transactions' input to create and submit their own profitable transactions.
While attackers are not required to use a bot contract to mount displacement attacks, using a smart contract allows them to limit their loss as they can abort the execution in case of an unexpected event.
However, detecting displacement attacks that directly interact with the contract that is susceptible to displacement is tremendously hard as there is no possible way to distinguish between an attacker and a benign user that just happened to send a transaction to the susceptible contract.
Our detection is therefore limited towards finding attackers that perform displacement attacks using bot contracts.
The general idea is to detect displacement by checking for every transaction $T$ if there exists a subsequent transaction $T'$ with a gas price lower than $T$ and a transaction index higher than $T$, where the input of $T'$ is contained inside the input of $T$.
However, detecting displacement in the wild can become quite challenging due to a large number of possible combinations.
A naive approach would be to obtain a list of every internal and external transaction per contract and then compare every transaction to every subsequent transaction. 
However, given that a single contract can have easily thousands of transactions, this approach would quickly result in a combinatorial explosion.
Moreover, obtaining internal transactions requires re-executing all past transactions which results in a significant amount of time given that the Ethereum blockchain currently has more than 1 billion transactions.
Our goal is therefore to focus only on external transactions and follow a more efficient approach that might sacrifice completeness but preserve soundness.
We begin by splitting the range of blocks that are to be analyzed into windows of 100 blocks and slide them with an offset of 20 blocks. 
This approach has the advantage that each window can be analyzed in parallel.
Inside each window, we iterate block by block, transaction by transaction, and split the input bytes of each transaction into $n$-grams of 4 bytes with an offset of 1 byte and check whether at least 95\% of the $n$-grams match with $n$-grams of previous transaction inputs. Since we focus on detecting displacement attacks performed via bot contracts, we cannot use 100\% matching, because the victim's external transaction will be encapsulated inside the attacker's external transaction along with some command-and-control data.
Each window has its own Bloom filter that memorizes previously observed $n$-grams.
A Bloom filter is a probabilistic data structure
that can quickly tell if an element has already been observed before or if it definitely has not been observed before. Thus, Bloom filters may yield false positives, but no false negatives.
The idea is first to use a Bloom filter to perform a quick probabilistic search and only perform an exhaustive linear search if the filter finds that at least 95\% of a transaction's $n$-grams are contained in the filter.
Our Bloom filters can hold up to $n$ = 1M elements with a false positive rate $p$ = 1\%, which according to Bloom \cite{bloom1970space}, requires having $k$ = 6 different hash functions:
\begin{equation}
m = -\frac{n \ln p}{(\ln 2)^{2}}
\end{equation}
\begin{equation}
k = \frac{m}{n} \ln 2
\end{equation}
We bootstrapped our 6 hash functions using the Murmur3 hash function as a basis.
The result of each hash function is an integer that acts as an index on the Bloom filter's bit array. The bit array is initialized at the beginning with zeros, and a value of one is set for each index returned by a hash function (see Figure~\ref{fig:displacement_detection}).
An $n$-gram is considered to be contained in the filter if all indices of the 6 hash functions are set to one.
We use interleaved $n$-grams because the input of a copied transaction might be included at any position in the attacker's input. 
Once our linear search finds two transactions $T_A$ and $T_V$ with matching inputs, we check whether the following three heuristics hold:
\begin{description}
    \item[Heuristic 1:] The sender of $T_A$ and $T_V$ as well as the receiver of $T_A$ and $T_V$ must be different. The receiver of $T_A$ and $T_V$ has to be different to make sure that we only detect displacement attacks that are performed by bot contracts.
    \item[Heuristic 2:] The gas price of $T_A$ must be larger than the gas price of $T_V$.
    \item[Heuristic 3:] We split the transaction input of $T_A$ and $T_V$ into sequences of 4 bytes, and the ratio between the number of the sequences must be at least 25\%.
    This heuristic requires that the byte sequences from $T_V$ conform with at least 25\% of the byte sequences of $T_A$ to avoid false positives. Without this restriction, it is very common for transactions with very small inputs to match by chance against transactions with very large inputs.
\end{description}

\noindent
However, the aforementioned heuristics may not filter out all the benign cases and therefore produce false positives. As a result, we filter out the benign cases by applying a runtime validation on the transaction inputs.
The heuristics are still useful and necessary since the validation process is computationally very intensive and the heuristics help us reduces the number of cases to validate and thus save time.
To validate that $T_A$ is a copy of $T_V$, we run in a simulated environment first $T_A$ before $T_V$ and then $T_V$ before $T_A$. We report a finding if the number of executed EVM instructions is different across both runs for $T_A$ and $T_V$, as this means that $T_A$ and $T_V$ influence each other. During our experiments, we noted, that some bot contracts included code that checks if the miner address of the block that is currently being executed is not equal to zero.
We think that the goal of this mechanism could be to prevent transactions from being run locally.

\noindent
\newline
\textbf{Limitations.} With more than 11 million blocks and over 1 billion transactions, we were compelled to make trade-offs between efficiency and completeness.
To be able to scan the entire blockchain for displacement attacks in a reasonable amount of time, we decided to set a window size of 100 blocks, meaning that we could not detect displacement attacks were an attacker's transaction and a victim's transaction are more than 100 blocks apart. 
Another limitation is that our heuristics only focus on detecting displacement attacks performed by bot contracts. For example, attackers can also send a transaction directly to the contract that is susceptible to displacement, without going through a bot contract. However, it is difficult for us to distinguish between benign users and attackers in such a case. Therefore, we decided to focus only on detecting bot contracts since a benign user would not use such a contract to perform a transaction to the susceptible contract.
Thus, our heuristics might produce false negatives and our results should be considered as a lower bound only.

\subsection{Detecting Insertion}

We limit our detection to insertion attacks on decentralized exchanges (DEXes).
At the time of writing, we are not aware of any other use case where insertion attacks are applied in the wild.
DEXes are decentralized platforms where users can trade their ERC-20 tokens for ether or other ERC-20 tokens via a smart contract.
Uniswap is currently the most popular DEX in terms of locked value with 3.15B USD locked\footnote{https://defipulse.com/}.
There exist two genres of DEXes, order book-based DEXes and automated market maker-based (AMM) DEXes. 
While order book-based DEXes match prices based on so-called 'bid' and 'ask' orders, AMM-based DEXes match and settle trades automatically on-chain via a smart contract, without the need of third party service.
AMMs are algorithmic agents that follow a deterministic approach to calculate the price of a token.
Uniswap, for example, is an AMM-based DEX, which computes for every trade the price of a token using the equation of a constant product market maker (CPMM):

\begin{equation}
[x] \times [y] = k
\end{equation}

\noindent
where $[x]$ is the current reserve of token $x$ and $[y]$ is the current reserve of token $y$.
Trades must not change the product $k$ of a pair's reserve. 
Thus, if the underlying token reserves decrease as a trader is buying, the token price increases. The same holds in the opposite direction: if the underlying token's reserve increases while a trader is selling, the token price decreases.
Despite being simple, CPMMs are incredibly susceptible to price slippage. 
Price slippage refers to the difference between a trade's expected price and the price at which the trade is executed.
Given the public nature of blockchains, attackers can observe large buy orders before miners pick them up by monitoring the transaction pool. 
These large buy orders will have a significant impact on the price of a token. 
Leveraging this knowledge and the fact that miners order transactions based on transaction fees, attackers can insert their buy order in front of an observed large buy order and insert a sell order after the observed large buy order to profit from the deterministic price calculation. 
\begin{figure}
  \centering
  \includegraphics[width=0.4\textwidth]{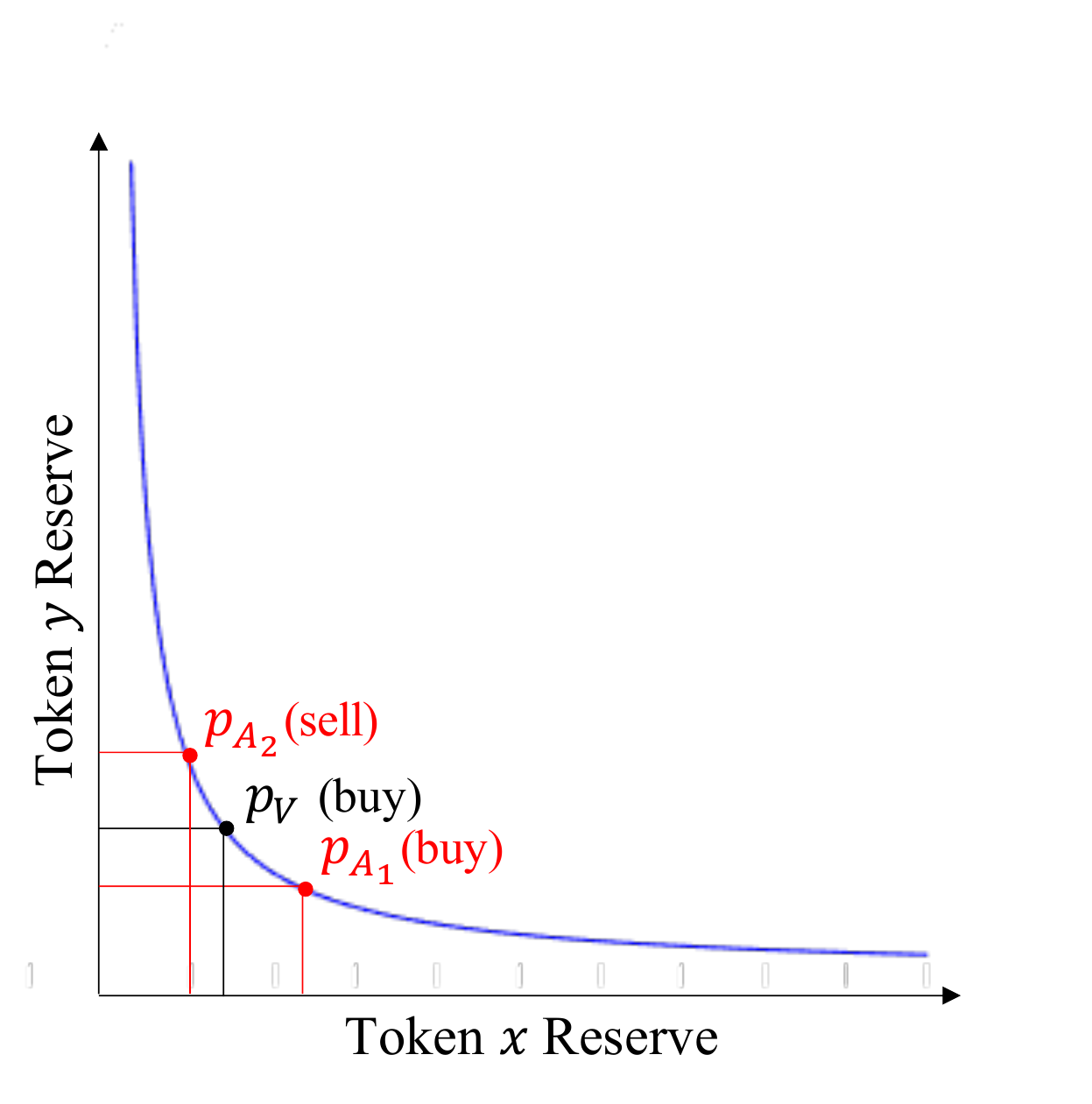}
  \caption{An illustrative example of an insertion attack on an AMM-based DEX that uses CPMM.}
  \label{fig:uniswap_insertion_attack}
\end{figure}
Figure~\ref{fig:uniswap_insertion_attack} depicts an example of an insertion attack on an AMM-based DEX that uses CPMM.
Let us assume that a victim $V$ wants to purchase some tokens at a price $p$. 
Let us also assume that an attacker $A$ observes $V$'s transaction and sends in two transactions: 1) a buy transaction which also tries to purchase some tokens at a price $p$, but with a gas price higher than $V$, and 2) a sell transaction that tries to sell the purchased tokens, but with a gas price lower than $V$. 
Since $A$ pays a higher gas price than $V$, $A$'s purchase transaction will be mined first and $A$ will be able to purchase the tokens at price $p$, where $p = p_{A_1}$ (\cf Figure~\ref{fig:uniswap_insertion_attack}). 
Afterwards, $V$'s transaction will be mined. However, $V$ will purchase tokens at a higher price $p_V$, where $p_V > p_{A_1}$ due to the imbalance in the token reserves (see Equation 3). 
Finally, $A$'s sell transaction will be mined, for which $A$ will sell its tokens at price $p_{A_2}$, where $p_{A_2} > p_{A_1}$ and therefore $A$ making profit.
Our detection algorithm exploits the fact that DEXes depend on the ERC-20 token standard.
The ERC-20 token standard defines many functions and events that enable users to trade their tokens between each other and across exchanges.
In particular, whenever a token is traded, a so-called \texttt{Transfer} event is triggered, and information about the sender, receiver, and the amount is logged on the blockchain.
We combine this information with transactional information (\eg transaction index, gas price, \etc) to detect insertion attacks.
We define a transfer event as $E = (s, r, a, c, h, i, g)$, where $s$ is the sender of the tokens, $r$ is the receiver of the tokens, $a$ is the number of transferred tokens, $c$ is the token's contract address, $h$ is the transaction hash, $i$ is the transaction index, and $g$ is the gas price of the transaction.
We detect insertion attacks by iterating block by block through all the transfer events and checking if there are three events $E_{A_1}$, $E_{V}$, and $E_{A_2}$ for which the following six heuristics hold:

\begin{description}
    \item[Heuristic 1:] The exchange transfers tokens to $A$ in $E_{A_1}$ and to $V$ in $E_{V}$, and the exchange receives tokens from $A$ in $E_{A_2}$. Moreover, $A$ transfers tokens in $E_{A_2}$ that it received previously in $E_{A_1}$.
g    Thus, the sender of $E_{A_1}$ must be identical to the sender of $E_{V}$ as well as the receiver of $E_{A_2}$, and the receiver of $E_{A_1}$ must be identical to the sender of $E_{A_2}$
    (\ie $s_{A_1} = s_{V} = r_{A_2} \wedge r_{A_1} = s_{A_2}$).
    \item[Heuristic 2:] The number of tokens bought by $E_{A_1}$ must be similar to the number of tokens sold by $E_{A_2}$. To avoid false positives, we set a conservative threshold of 1\%.
    Hence, the difference between token amount $a_{A_1}$ of $E_{A_1}$ and token amount $a_{A_2}$ of $E_{A_2}$ cannot be more than 1\% (\ie $\frac{|a_{A_1} - a_{A_2}|}{max(a_{A_1}, a_{A_1})} \leq 0.01$).
    \item[Heuristic 3:] The token contract addresses of $E_{A_1}$, $E_{V}$, and $E_{A_2}$ must be identical (\ie $c_{A_1} = c_{V} = c_{A_2}$).
    \item[Heuristic 4:] The transaction hashes of $E_{A_1}$, $E_{V}$, and $E_{A_2}$ must be dissimilar (\ie $h_{A_1} \neq h_{V} \neq h_{A_2}$).
    \item[Heuristic 5:] The transaction index of $E_{A_1}$ must be smaller than the transaction index of $E_{V}$, and the transaction index of $E_{V}$ must be smaller than the transaction index of $E_{A_2}$ (\ie $i_{A_1} < i_{V} < i_{A_2}$).
    \item[Heuristic 6:] The gas price of $E_{A_1}$ must be larger than the gas price of $E_{V}$, and the gas price of $E_{A_2}$ must be less of equal to the gas price of $E_{V}$ (\ie $g_{A_1} > g_{V} \geq g_{A_2}$).
\end{description}

\noindent
\textbf{Limitations.}
Our heuristics assume that insertion attacks always occur within the same block. 
This assumption enables us to check blocks in parallel since we only need to compare transactions within a block.
However, this assumption does not always hold in reality, as transactions might be scattered across different blocks during the mining process. 
Thus, there might exist insertion attacks that were performed across multiple blocks, which our heuristics do not detect and therefore might result in false negatives.

\subsection{Detecting Suppression}

In suppression, an attacker's goal is to withhold a victim's transaction by submitting transactions to the network that consume large amounts of gas and fill up the block gas limit such that the victim's transaction cannot be included anymore.
There are several ways to achieve this. The naive approach uses a smart contract that repeatedly executes a sequence of instructions in a loop to consume gas. This strategy can either be controlled or uncontrolled.
In a controlled setting, the attacker repeatedly checks how much gas is still left and exits the loop right before all gas is consumed such that no out-of-gas exception is raised.
In an uncontrolled setting, the attacker does not repeatedly check how much gas is left and lets the loop run until no more gas is left and an out-of-gas exception is raised.
The former strategy does not consume all the gas and does not raise an exception which makes it less obtrusive, while the latter strategy does consume all the gas but raises an exception which makes it more obtrusive.
However, a third strategy achieves precisely the same result without running code in an infinite loop.
If we think about it, the attacker's goal is not to execute useless instructions but rather to force miners to consume the attacker's gas units to fill up the block.
The EVM proposes two ways to raise an error during execution, either through a revert or an assert. The difference between revert and assert is that the former returns the unused gas to the transaction sender, while the latter consumes the entire gas limit initially specified by the transaction sender.
Hence, an attacker can exploit this and call an assert to consume all the provided gas with just one instruction.
Our goal is to detect transactions that employ one of the three aforementioned suppression strategies: \emph{controlled gas loop}, \emph{uncontrolled gas loop}, and \emph{assert}.
We start by clustering for each block all transactions with the same receiver, as we assume that attackers send multiple suppression transactions to the same bot contract. Afterwards, we check the following heuristics for each cluster:
\begin{description}
    \item[Heuristic 1:] The number of transactions within a cluster must be larger than one.
    \item[Heuristic 2:] All transactions within a cluster must have consumed more than 21,000 gas units. The goal of this heuristic is to filter out transactions that only transfer value (\ie ether), but do not execute code.
    \item[Heuristic 3:] The ratio between gas used and gas limit must be larger than 99\% for all transactions within the cluster.
\end{description}

\noindent
If we happen to find a cluster that fulfills the heuristics mentioned above, we check whether at least one of the neighbouring blocks (\ie the previous block and the subsequent block) also contains a cluster that satisfies the same heuristics. 
We assume that an attacker tries to suppress transactions for a sequence of blocks. 
Finally, we try to detect if an attacker employs one of three suppression strategies by retrieving and analyzing the execution trace of the first transaction in the cluster. An execution trace consists of a sequence of executed instructions.
We detect the first strategy by checking if the transaction did not raise an exception and if the instruction sequence \texttt{[GAS, GT, ISZERO, JUMPI]} is executed more than ten times in a loop. 
This particular instruction sequence checks how much gas is left and jumps towards a different code location, if the amount of gas is lower than a given value.
We detect the second strategy by checking if the transaction raised an exception via a revert and if the instruction sequence \texttt{[SLOAD, TIMESTAMP, ADD, SSTORE]} is executed more than ten times in a loop. 
This particular instruction sequence  increments a persistent counter residing in storage with the current timestamp in order to consume a large amount of gas.
Finally, we detect the third strategy by checking if the transaction raised an exception via an assert.

\noindent
\newline
\textbf{Limitations.} Our heuristics follow two major assumptions.
First, we assume that an attacker always sends multiple transactions to the same bot contract. However, an attacker could also just send one transaction and deploy multiple bot contracts for single use. 
Second, we assume that an attacker always tries to suppress more than just one block. However, an attacker could also just try to suppress one block.
While in practice we always observed that attackers tried to suppress multiple blocks and sent multiple transactions as well as reused the same bot contract, it is still possible that some attackers do not follow this pattern and therefore our heuristics might produce false negatives.

\section{Analyzing Frontrunning Attacks}

In this section, we analyze the results of our large scale measurement study on detecting frontrunning in Ethereum.

\subsection{Experimental Setup}

We implemented our detection modules using Python with roughly 1,700 lines of code\footnote{Code and data are publicly available on GitHub: \url{https://github.com/christoftorres/Frontrunner-Jones}.}
We run our modules on the first 11,300,000 blocks of the Ethereum blockchain, ranging from July 30, 2015 to November 21, 2020. 
All our experiments were conducted using a machine with 128 GB of memory and 10 Intel(R) Xeon(TM) L5640 CPUs with 12 cores each and clocked at 2.26 GHz, running 64 bit Ubuntu 16.04.6 LTS.

\subsection{Validation}
Since our work is the first to systematically study the three different types of frontrunning by leveraging historical blockchain data on such a large scale, we are missing a ground truth against which we can compare our results. Our goal was therefore to design very precise and rather conservative heuristics that might yield false negatives, but no false positives. We started with a rather liberal definition of our heuristics and did several iterations, where we regularly checked for outliers and tried to tighten the heuristics after each iteration whenever we discovered false positives in our preliminary results. 
After finding no more false positives we ran our final experiments, which resulted in over 200K transactions being labeled as either displacement, insertion, or suppression frontrunning attacks. Since checking all of these 200K transactions manually is extremely cumbersome, we decided to select a random sample of 100 findings for each type of frontrunning attack and manually check them for false positives. For displacement, we tried to reverse engineer the code of the identified bot contract to see if the code was proxying the transaction input to a specified contract destination. For insertion, we checked if the two reported attacker transactions and the whale transaction were indeed buying or selling the exact same token via the same exchange. Finally, for suppression, we tried to reverse engineer the reported bot contract and to check if the contract would probe who is the last purchaser of a ticket of a specific lottery or gambling contract and try to consume the entire gas in case the last purchaser was a specific address. Following these steps, our manual validation did not reveal any false positives. However, as already mentioned previously, our heuristics have some limitations which might result in false negatives. Hence, all the results presented in this paper should be interpreted only as lower bounds, and they might only show the tip of the iceberg.

\subsection{Analyzing Displacement}

\begin{figure*}
    \begin{center}
        \includegraphics[width=1.0\textwidth]{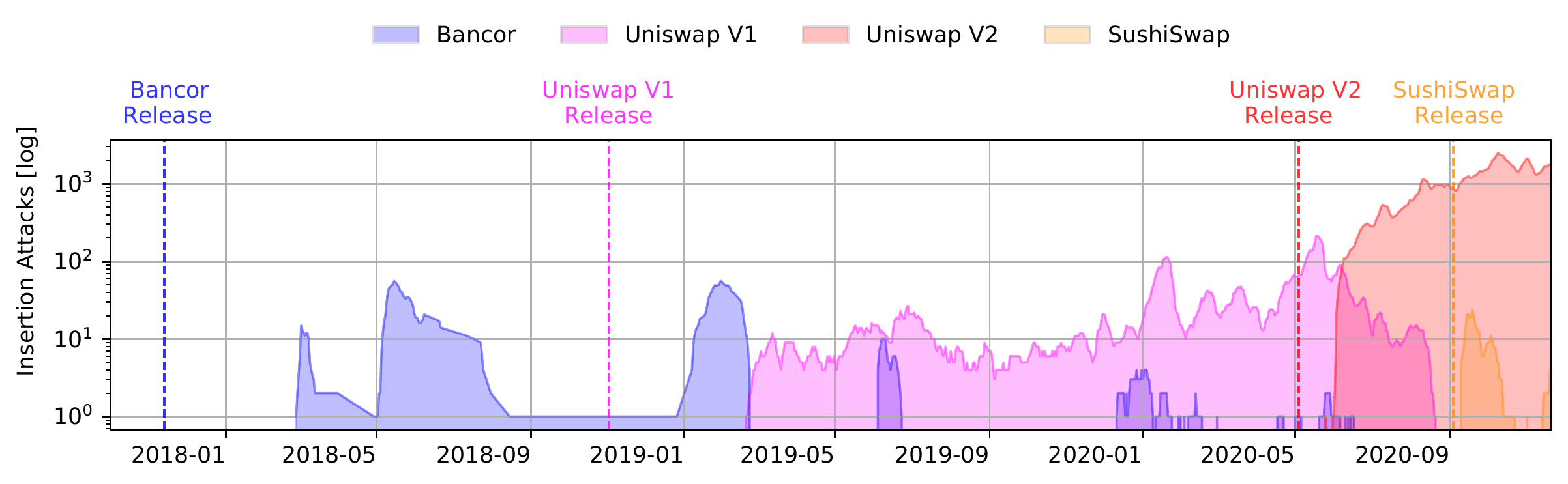}
    \end{center}
    \caption{\label{fig:insertion_attacks} Weekly average of daily insertion attacks per decentralized exchange.}
\end{figure*}

\noindent
\textbf{Overall Results.}
We identified a total of 2,983 displacement attacks from 49 unique attacker accounts and 25 unique bot contracts.
Using the graph analysis defined in Section~\ref{sec:identifying-attackers} we identified 17 unique attacker clusters.

\noindent
\newline
\textbf{Profitability.}
We compute the gain of an attacker $A$ on each displacement attack by searching how much ether $EOA_A$ receives among the internal transactions triggered by $T_A$.
Additionally, we obtain the profit by subtracting the attack cost from the gain, where cost is defined solely by the fees of $T_A$.
Finally, for each attack we convert the ether cost and profit into USD by taking the conversion rate valid at the time of the attack.

\noindent
\newline
\textbf{Attacks.} We can see in Table~\ref{tbl:displacement-attack-distributions} the distribution of each variable we collected per displacement attack.
The cost and the profit do not appear to be very high for most of the attacks, but the distributions of both variables present very long tails to the right.
Additionally, we compute the Gas Price $\Delta$ as the gas price of $T_A$ minus the gas price of $T_V$.
This value indicates how much the attacker $A$ is willing to pay to the miners so they execute $T_A$ before $T_V$.
Table~\ref{tbl:displacement-attack-distributions} shows that most of the attacks contain a very small gas price difference in GWei (and cannot be represented with only two digits of precision), but there are very extreme cases with a difference close to 50 GWei.
Furthermore, we compute the Block $\Delta$ to indicate how many blocks are between the execution of $T_A$ and $T_V$.
Again we can see in Table~\ref{tbl:displacement-attack-distributions} that for most of the attacks, both transactions were executed in the same block, but there are some extreme cases with a long block distance of 19 blocks.

\begin{table}[H]
  \centering
  \begin{adjustbox}{width=\columnwidth,center}
  \begin{tabular}{l r r r r}
\toprule
{} &     Cost (USD) &     Profit (USD) & Gas Price $\Delta$ (GWei) & Block $\Delta$ \\
\midrule
mean  &    14.28 &   1,537.99 &            0.43 &        0.78 \\
std   &    18.25 &   7,162.80 &            2.65 &        2.37 \\
min   &     0.01 &       0.00 &            0.00 &        0.00 \\
25\%   &     4.36 &       1.14 &            0.00 &        0.00 \\
50\%   &     9.48 &     158.53 &            0.00 &        0.00 \\
75\%   &    16.64 &     851.04 &            0.00 &        0.00 \\
max   &   311.69 & 223,150.01 &           52.90 &       19.00 \\
\bottomrule
  \end{tabular}
  \end{adjustbox}
  \caption{Distributions for displacement attacks.}
  \label{tbl:displacement-attack-distributions}
\end{table}

\noindent
\newline
\textbf{Attacker Clusters.}
Each of the 17 identified clusters contains bot accounts with different bytecode, with the exception of one cluster that contains three bot accounts with the exact same bytecode.
Table~\ref{tbl:displacement-cluster-distributions} presents the distribution of each attacker cluster variable.
The first variable describes profit, where we can see that a single attacker mounted 2,249 attacks making an accumulated profit of more than 4.1M USD while spending over 40K USD in transaction fees.
We can also see that the attacker used 16 different accounts and 3 different bots to mount its attacks.
The minimum amount of profit that an attacker made with displacement is 0.01 USD.
Overall, the average number of attacks per attacker cluster is 175.47 attacks, using 2.88 accounts and 1.47 bots. However, we also observe from the distribution that at least half of the attackers only use one account and one bot contract.

\begin{table}[H]
  \centering
  \begin{adjustbox}{width=\columnwidth,center}
  \begin{tabular}{l r r r r r}
\toprule
{} &      Cost (USD) &       Profit (USD) & Attacks & Attacker Accounts & Bot Contracts \\
\midrule
mean  &  2,505.09 &   269,872.45 &  175.47 &              2.88 &          1.47 \\
std   &  9,776.51 & 1,005,283.40 &  555.03 &              3.89 &          0.80 \\
min   &      0.05 &         0.01 &    1.00 &              1.00 &          1.00 \\
25\%   &      0.14 &         3.53 &    1.00 &              1.00 &          1.00 \\
50\%   &      3.98 &       726.70 &    5.00 &              1.00 &          1.00 \\
75\%   &     65.78 &     4,670.94 &    8.00 &              3.00 &          2.00 \\
max   & 40,420.63 & 4,152,270.01 & 2249.00 &             16.00 &          3.00 \\
\bottomrule
  \end{tabular}
  \end{adjustbox}
  \caption{Distributions for displacement attacker clusters.}
  \label{tbl:displacement-cluster-distributions}
\end{table}

\subsection{Analyzing Insertion}

\textbf{Overall Results.}
We identified a total of 196,691 insertion attacks from 1,504 unique attacker accounts and 471 unique bot contracts.
Using the graph analysis defined in Section~\ref{sec:identifying-attackers} we identified 98 unique attacker clusters.

\noindent
\newline
\textbf{Profitability.}
We compute the cost for each attack as the sum of the amount of ether an attacker spent in $T_{A_1}$ and the fees imposed by transactions $T_{A_1}$ and $T_{A_2}$.
Additionally, we compute the profitability of an attack as the amount of ether an attacker gained in $T_{A_2}$ minus the cost.
Finally, for each attack we convert the ether cost and profit into USD by taking the conversion rate valid at the time of the attack.

\noindent
\newline
\textbf{Attacks.} We can see in Table~\ref{tbl:insertion-attack-distributions} the distribution of each variable we collected per insertion attack.
The cost and the profit do not appear to be very high for most of the attacks, but the distributions of both variables present very long tails to the right.
Note that the profit also present very large negative values to the left, meaning that there are extreme cases of attackers losing money.
Additionally, we compute the Gas Price $\Delta_1$ and Gas Price $\Delta_2$ as the gas price of $T_{A_1}$ minus the gas price of $T_V$, and the gas price of $T_V$ minus the gas price of $T_{A_2}$ respectively.
This value indicates how much the attacker $A$ is willing to pay to the miners so they execute $T_{A_1}$ before $T_V$ and also if $T_{A_2}$ can be executed after $T_V$.
Table~\ref{tbl:insertion-attack-distributions} shows that 25\% of the attacks contain a very small Gas Price $\Delta_1$ in GWei (and cannot be represented with only two digits of precision), but that half or more paid a significant difference, reaching some extreme cases of more than 76K GWei.
For Gas Price $\Delta_2$ most of the attacks have a very small value, but there are extreme cases, which mean that some attacks are targeting transactions with very high gas prices.

\begin{table}[H]
  \centering
  \begin{adjustbox}{width=\columnwidth,center}
  \begin{tabular}{l r r r r}
\toprule
{} &       Cost (USD) &     Profit (USD) & Gas Price $\Delta_1$ (GWei) & Gas Price $\Delta_2$ (GWei)  \\
\midrule
mean  &      19.41 &      65.05 &                      407.63 &                         3.88 \\
std   &      51.15 &     233.44 &                    1,897.47 &                       137.12 \\
min   &       0.01 & -10,620.61 &                        0.00 &                         0.00 \\
25\%   &       4.09 &       7.86 &                        0.00 &                         0.00 \\
50\%   &       7.74 &      24.07 &                        5.25 &                         0.00 \\
75\%   &      15.23 &      62.92 &                       74.10 &                         0.00 \\
max   &   1,822.22 &  20,084.01 &                   76,236.09 &                    27,396.63 \\
\bottomrule
  \end{tabular}
  \end{adjustbox}
  \caption{Distributions for insertion attacks.}
  \label{tbl:insertion-attack-distributions}
\end{table}

\begin{figure*}
     \begin{center}
         \includegraphics[width=1.0\textwidth]{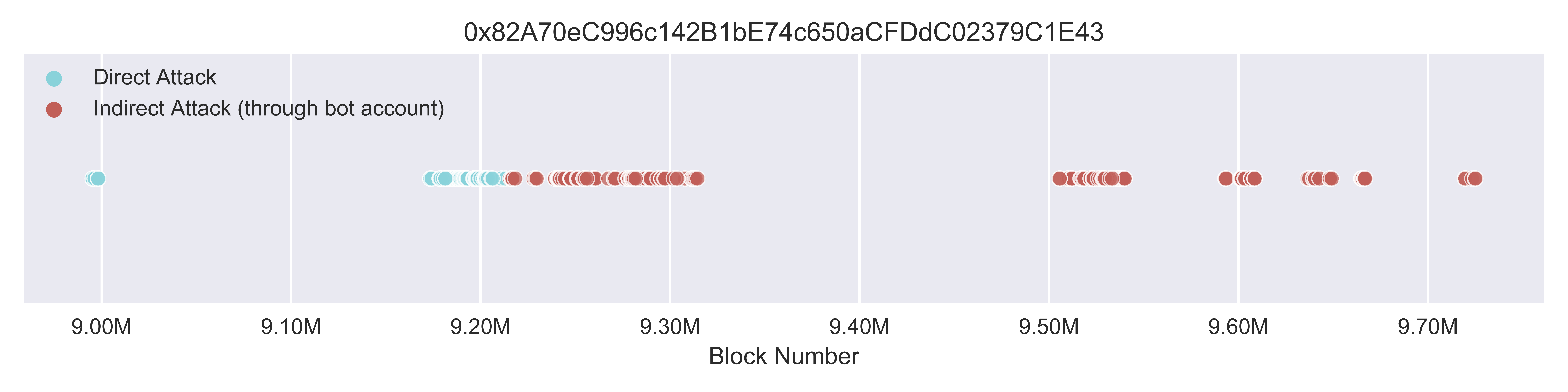}
         \includegraphics[width=1.0\textwidth]{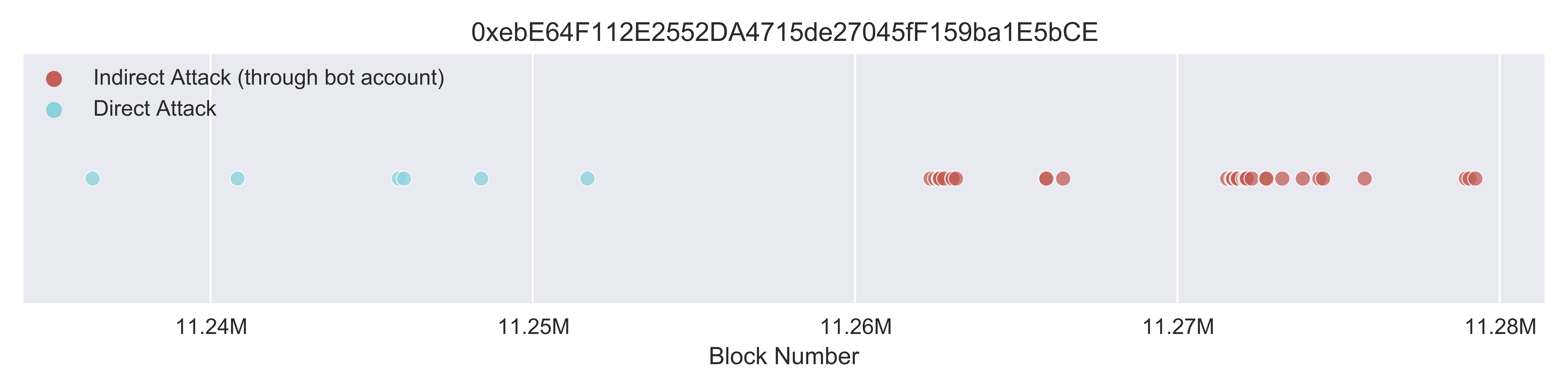}
     \end{center}
     \caption{\label{fig:insertion-interface} Two examples of attackers changing their strategies over time from direct attacks (\ie using directly an exchange) to indirect attacks (\ie using a bot contract).}
\end{figure*}

\noindent
\newline
\textbf{Gas Tokens.}
We analyzed how many attacks were mounted using gas tokens. Gas tokens allow attackers to reduce their gas costs. 
We found that 63,274 (32,17\%) of the insertion attacks we measured were performed using gas tokens.
48,281 (76.3\%) attacks were mounted using gas tokens only for the first transaction $T_{A_1}$, 1,404 (2.22\%) attacks were mounted by employing gas tokens only for the second transaction $T_{A_2}$, and 13,589 (21.48\%) attacks were mounted by employing gas tokens for both transactions $T_{A_1}$ and $T_{A_2}$. 
We also found that 24,042 (38\%) of the attacks used GST2, 14,932 (23.6\%) used ChiToken, and 24,300 (38.4\%) used their own implementation or copy of GST2 and ChiToken.

\noindent
\newline
\textbf{Exchanges and Tokens.}
We identified insertion attacks across 3,200 different tokens on four exchanges: Bancor, Uniswap V1, Uniswap V2, and SushiSwap.
Figure~\ref{fig:insertion_attacks} depicts the weekly average of daily insertion attacks per exchange.
The first AMM-based DEX to be released on Ethereum was Bancor in November 2017. 
We observe from Figure~\ref{fig:insertion_attacks} that the first insertion attacks started in February 2018, targeting the Bancor exchange. We also see that the number of insertion attacks increased tremendously with the rise of other DEXes, such as Uniswap V1 and Uniswap V2.
While it took 3 months for attackers to launch their first insertion attacks on Uniswap V1, it only took 2 weeks to launch attacks on Uniswap V2 and 5 days to launch attacks on SushiSwap.
This is probably due to the core functionality of Uniswap V1 and Uniswap V2 being the same and that SushiSwap is a direct fork of Uniswap V2.
Thus, for attackers it was probably straightforward to take their existing code for Uniswap V1 and adapt it to attack Uniswap V2 as well as SushiSwap.
The peak of insertion attacks was on October 5, 2020, with 2,749 daily attacks.
We measured in total 3,004 attacks on Bancor, 13,051 attacks on Uniswap V1, 180,185 attacks on Uniswap V2, and 451 attacks on SushiSwap.
Table~\ref{tbl:intertion-exchanges} shows the different combinations of exchanges that attackers try to frontrun.
We see that most of the attackers focus on attacking Uniswap V2, with 72 attacker clusters (73.47\%).
We also see that 92.86\% of the attackers only focus on attacking one exchange.
Moreover, we observed one attacker that attacked all the 4 exchanges, 2 attackers that attacked Uniswap V1 and Uniswap V2, and 4 attackers that attacked Uniswap V2 and SushiSwap. The latter is expected since SushiSwap is a direct fork of Uniswap V2. Hence, the attackers can reuse their code from Uniswap V2 to attack SushiSwap. What is interesting though, is the fact that no attacker is attacking only SushiSwap, we see that attacker always attack SushiSwap in conjunction to another exchange. 

\begin{table}
 \centering
 \begin{adjustbox}{width=0.95\columnwidth,center}
\begin{tabular}{lr}
\toprule
Exchange Combination & Attacker Clusters \\
\midrule
Uniswap V2                                &        72 \\
Uniswap V1                                &        16 \\
SushiSwap, Uniswap V2                     &         4 \\
Bancor                                    &         3 \\
Uniswap V1, Uniswap V2                    &         2 \\
Bancor, SushiSwap, Uniswap V1, Uniswap V2 &         1 \\
\bottomrule
\end{tabular}
  \end{adjustbox}
 \caption{Exchange combination count by attacker cluster.}
 \label{tbl:intertion-exchanges}
\end{table}

\noindent
\newline
\textbf{Attack Strategies.}
In 186,960 cases (95.05\%) the attackers sold the exact same amount of tokens that they purchased.
Thus, an easy way to spot insertion attacks on decentralized exchanges, could be to check for two transactions that have the same sender and receiver, and where the first transaction buys the same amount of tokens that the second transaction sells.
However, some attackers try to obscure their buy and sell transactions by using different sender accounts. We found 86,038 cases of attacks (43.74\%) where attackers used a different sender address to buy tokens than to sell tokens.
Moreover, besides trying to hide their sender accounts, attackers also try to hide in some cases the addresses of their bot contracts by using proxy contracts to forward for instance the call to buy tokens to the bot contracts. To the outsider it will look like two transactions with different receivers. We found only 5,467 cases (2.78\%) where the attackers are using proxy contracts to disguise calls to their bot contracts.
Insertion is the only attack type for which our heuristics can detect attacks that do not employ bot contracts.
For these cases, the attacker accounts call the DEXes directly.
From all the insertion attacks we detected, only 2,673 cases (0.01\%) fall in this category of direct attacks.
We included these attacks in most of the results, but we do not count them for the cluster computation since we cannot link the corresponding attacker accounts to any bot contract.
Figure~\ref{fig:insertion-interface} highlights examples of two accounts that changed their attack strategy over time. The attackers initially performed their attacks by calling directly the smart contract of exchanges, but then switched to bot contracts over time.
\newline
\newline
\noindent
\textbf{Attacker Clusters.}
Among the 98 attacker clusters that we identified, many of the bot contracts share the same bytecode.
The most extreme case is an attacker cluster that contains 80 bot contracts and all of them have the same bytecode.
We find that attackers were already able to make an accumulated profit of over 13.9M USD. 
From Table~\ref{tbl:insertion-cluster-distributions}, we see that an attacker makes on average a profit of over 130K USD per attacker cluster.
Moreover, the average profit per attack is 78.72 USD, whereas the median profit is 28.80 USD. 
The largest profit that has been made with a single attack was 20,084.01 USD.
However, not all the attacks were successful in terms of profit. 
We count 19,828 (10.08\%) attacks that resulted in an accumulated loss of roughly 1.1M USD. The largest loss that we measured was 10,620.61 USD. The average loss is 56.93 USD per attack and the median loss is 14.26 USD per attack. Thus, the average loss is still lower than the average profit, meaning that insertion attacks are profitable despite bearing some risks.

\begin{table}[t]
  \centering
  \begin{adjustbox}{width=\columnwidth,center}
  \begin{tabular}{l r r r r r}
\toprule
{} &      Cost (USD) &       Profit (USD) & Attacks & Attacker Accounts & Bot Contracts \\
\midrule
mean  &  38,807.63 &   130,246.93 &  1979.78 &             14.87 &          4.81 \\
std   & 135,352.00 &   462,464.36 &  6053.68 &             90.59 &         10.09 \\
min   &       0.98 &    -2,319.42 &     1.00 &              1.00 &          1.00 \\
25\%   &      43.84 &        -9.78 &     4.25 &              1.00 &          1.00 \\
50\%   &     419.74 &       691.48 &    68.50 &              2.00 &          2.00 \\
75\%   &   3,510.94 &     8,350.46 &   529.25 &              3.00 &          4.00 \\
max   & 686,850.37 & 2,262,411.95 & 39162.00 &            891.00 &         80.00 \\
\bottomrule
  \end{tabular}
  \end{adjustbox}
  \caption{Distributions for insertion attacker clusters.}
  \label{tbl:insertion-cluster-distributions}
\end{table}

\noindent
\newline
\textbf{Competition.}
We found among our detected results 5,715 groups of at least two insertion attacks that share the same block number, victim transaction and exchanged token but with different attackers.
Included in those groups, we found 270 cases where at least two of the attackers targeting the same victim belong to the same attacker cluster.
To explain this phenomenon, we have three hypothesis.
The first one is that an attacker would not interfere with its own attacks, hence, our attacker clustering mechanism is incorrect.
Since our methodology is based on heuristics and we have no ground truth to validate them, we could expect to find occasional errors.
However, since the heuristics are simple and reasonable enough, we also consider the next two hypothesis.
The second one is that some attackers might not be clever enough to coordinate multiple agents working in parallel, and the self-interference could be an accident.
And third, the parallel attacks could be a tactic to split the movements of funds into smaller amounts to avoid becoming the target of other attackers.
For example, we found two instances where attackers became victims at the same time, namely accounts 0x5e334032Fca55814dDb77379D8f99c6eb30dEa6a and 0xB5AD1C4305828636F32B04E5B5Db558de447eAff in blocks 11,190,219 and 11,269,029, respectively.

\subsection{Analyzing Suppression}

\begin{table*}
  \centering
  \begin{adjustbox}{width=\textwidth,center}
  \begin{tabular}{ccrrrrrr}
\toprule
Suppressed Contract Address & Contract Name & Attacks & Rounds & Transactions &  Attackers &  Bot Contracts & Attacker Clusters \\
\midrule
 \href{https://etherscan.io/address/0xDd9fd6b6F8f7ea932997992bbE67EabB3e316f3C}{0xDd9fd6b6F8f7ea932997992bbE67EabB3e316f3C} &   Last Winner &      16 &     20 &          304 &         27 &     5 &                 2 \\
 \href{https://etherscan.io/address/0xA62142888ABa8370742bE823c1782D17A0389Da1}{0xA62142888ABa8370742bE823c1782D17A0389Da1} &    FoMo3Dlong &      12 &    188 &         5875 &         81 &     8 &                 4 \\
 \href{https://etherscan.io/address/0x5D0d76787D9d564061dD23f8209F804a3b8AD2F2}{0x5D0d76787D9d564061dD23f8209F804a3b8AD2F2} &    Peach Will &       6 &     52 &         1105 &         26 &     5 &                 2 \\
 \href{https://etherscan.io/address/0x2c58B11405a6a8154FD3bbC4CcAa43924f2BE769}{0x2c58B11405a6a8154FD3bbC4CcAa43924f2BE769} &           ERD &       3 &      3 &          207 &         20 &     2 &                 1 \\
 \href{https://etherscan.io/address/0x42CeaD70158235a6ca4868F3CFAF600c7A7b0ebB}{0x42CeaD70158235a6ca4868F3CFAF600c7A7b0ebB} &       ETH CAT &       2 &     23 &          929 &         20 &     2 &                 1 \\
 \href{https://etherscan.io/address/0xB7C2e4047Fb76508D4137BE787DaF28B013F00E6}{0xB7C2e4047Fb76508D4137BE787DaF28B013F00E6} &   Escape plan &       2 &      3 &           67 &         20 &     2 &                 1 \\
 \href{https://etherscan.io/address/0x29488e24cFdAA52a0b837217926C0c0853Db7962}{0x29488e24cFdAA52a0b837217926C0c0853Db7962} &     SuperCard &       1 &     25 &          319 &         17 &     1 &                 1 \\
 \href{https://etherscan.io/address/0xB4a448387403554616eB5B50aa4C48f75243a015}{0xB4a448387403554616eB5B50aa4C48f75243a015} &    Mobius2Dv2 &       1 &      4 &           82 &         19 &     1 &                 1 \\
 \href{https://etherscan.io/address/0x3e22bB2279d6Bea3Cfe57f3Ed608fC3B1DeaDADf}{0x3e22bB2279d6Bea3Cfe57f3Ed608fC3B1DeaDADf} &    Star3Dlong &       1 &      3 &           66 &          6 &     1 &                 1 \\
 \href{https://etherscan.io/address/0xD15E559f6BD5C785Db35E550F9FbC80045b0a049}{0xD15E559f6BD5C785Db35E550F9FbC80045b0a049} &           FDC &       1 &      3 &           44 &         18 &     1 &                 1 \\
 \href{https://etherscan.io/address/0x9954fF17909893B443E2EE825066373960c2735A}{0x9954fF17909893B443E2EE825066373960c2735A} &        F3DPRO &       1 &      1 &           41 &         18 &     1 &                 1 \\
 \href{https://etherscan.io/address/0xC75506dEAe7c01F47BCd330B324226CE9ba78e30}{0xC75506dEAe7c01F47BCd330B324226CE9ba78e30} &        FomoXP &       1 &      3 &           39 &         19 &     1 &                 1 \\
 \href{https://etherscan.io/address/0x0fe2247a20E779a879c647D2b9deA1b896FC0ccf}{0x0fe2247a20E779a879c647D2b9deA1b896FC0ccf} &           EFS &       1 &      1 &           33 &         16 &     1 &                 1 \\
 \href{https://etherscan.io/address/0xbAbED6ca5C86B2347D374e88251Ca8007C417f55}{0xbAbED6ca5C86B2347D374e88251Ca8007C417f55} &    The rabbit &       1 &      1 &           15 &         13 &     1 &                 1 \\
 \href{https://etherscan.io/address/0xb178EA2c9023bb2DD500a607505D2aa121F92A35}{0xb178EA2c9023bb2DD500a607505D2aa121F92A35} &       RichKey &       1 &      1 &            9 &          8 &     1 &                 1 \\
\bottomrule
\end{tabular}
  \end{adjustbox}
  \caption{List of contracts that were victims of suppression attacks.}
  \label{tbl:suppressed_contracts}
\end{table*}

\textbf{Overall Results.}
We identified a total of 50 suppression attacks originated from 98 attacker accounts and 30 bot contracts.
From these entities, we identified 5 unique attacker clusters using the graph analysis defined in Section~\ref{sec:identifying-attackers}.

\noindent
\newline
\textbf{Rounds, Success, and Failure.}
In this section we define a suppression attack as a sequence of rounds.
Each round starts with an investment transaction that sends ether to the victim's contract, which is added to a prize pool.
The round then continues with a sequence of one or more stuffing transactions.
When another participant interrupts the stuffing sequence by sending a new investment transaction, the participant becomes the new potential winner of the prize pool.
This event terminates the round in a failure state, because the attacker cannot claim the prize anymore.
Otherwise, if an interruption never occurs and the attacker can eventually claim the competition prize, the round is terminated with a success status.
Thus, we define the status of an entire suppression attack as the status of the last round in the corresponding sequence of rounds.
From the 50 suppression attacks we identified, 13 were successful and 37 failed.

\noindent
\newline
\textbf{Suppression Strategies.}
In Table~\ref{tbl:suppression_strategies} we show the distribution of suppression strategies split by successful and failed attacks.
We see that although the assert strategy is the most popular one, it is not the most successful one. The controlled gas loop strategy seems to be the most successful in terms of attacks.

\begin{table}[H]
  \centering
  \begin{adjustbox}{width=0.85\columnwidth,center}
  \begin{tabular}{l r r r}
    \toprule
    Suppression Strategy & Attacks & Successful & Failed \\
    \midrule
    Assert                  & 20 & 2 & 18 \\
    Controlled Gas Loop     & 18 & 8 & 10 \\
    Uncontrolled Gas Loop   & 12 & 3 &  9 \\
    \bottomrule
  \end{tabular}
  \end{adjustbox}
  \caption{Suppression strategies.}
  \label{tbl:suppression_strategies}
\end{table}

\noindent
\textbf{Profitability.}
In a suppression attack, the profit of the attacker $A$ is defined by the accumulated ether in the prize pool of the suppressed contract.
Note that the attack only obtains the prize if it succeeds.
Additionally, we subtract from the profit the attack cost which is defined by the sum of the initial investment on each round, and the accumulated fees of all the related transactions $T_{A_i}$.
Finally, for each attack we convert the ether cost and profit into USD by taking the conversion rate valid at the time of the attack.

\noindent
\newline
\textbf{Attacks.} We can see in Table~\ref{tbl:suppression-attack-distributions} the distribution of each variable we collected per suppression attack.
An interesting result is that at least 75\% of the attacks generate big losses.
However, there are also extreme cases with huge profits.
Hence, we could say that the suppression attacks are very risky but that they can also yield high rewards.
Along with the price and cost, we also count the number of rounds, blocks and transactions every attack contains.
We can observe, as expected in Table~\ref{tbl:suppression-attack-distributions}, how all these metrics grow together with the cost.
A suppression attack lasts on average 6.62 rounds and an attacker stuffs on average 29.70 blocks with an average of 182.70 transactions.

\begin{table}[H]
  \centering
  \begin{adjustbox}{width=\columnwidth,center}
  \begin{tabular}{l r r r r r}
\toprule
{} &      Cost (USD) &     Profit (USD) &     Rounds &    Blocks &  Transactions \\
\midrule
mean  &  2,349.65 &  20,725.24 &   6.62 &  29.70 &       182.70 \\
std   &  3,331.21 & 113,598.58 &  12.86 &  50.77 &       456.91 \\
min   &      4.67 & -10,741.12 &   1.00 &   2.00 &         6.00 \\
25\%   &    221.87 &  -1,893.26 &   1.00 &   4.00 &        12.50 \\
50\%   &    896.68 &    -284.81 &   2.00 &  10.00 &        33.50 \\
75\%   &  2,719.69 &     -14.93 &   4.75 &  21.50 &        88.75 \\
max   & 10,741.12 & 791,211.86 &  66.00 & 233.00 &     2,664.00 \\
\bottomrule
  \end{tabular}
  \end{adjustbox}
  \caption{Distributions for suppression attacks.}
  \label{tbl:suppression-attack-distributions}
\end{table}

\noindent
\textbf{Attacker Clusters.}
We identified 5 attacker clusters.
Among the attacker clusters, we found only two pairs of bot contracts sharing the same bytecode.
From Table~\ref{tbl:suppression-cluster-distributions}, we can see that the average profit per attacker cluster is 207,252.36 USD and that the largest profit made by an attacker cluster is over 777K USD.
However, we also see that mounting suppression attacks is expensive with an average of 23,496.52 USD, but still profitable with an average profit of 207,252.36 USD.
Also, we find that attackers mount on average 10 attacks and use on average around 19 attacker accounts and 6 bot contracts. 
There is one case where an attacker was responsible for mounting 18 different attacks using 42 different accounts and 14 different bots.

\begin{table}[H]
  \centering
  \begin{adjustbox}{width=\columnwidth,center}
  \begin{tabular}{l r r r r r}
\toprule
{} &      Cost (USD) &       Profit (USD) & Attacks & Attacker Accounts & Bot Contracts \\
\midrule
mean  & 23,496.52 & 207,252.36 &   10.00 &             19.60 &          6.00 \\
std   & 20,520.87 & 323,613.48 &    7.65 &             13.67 &          5.24 \\
min   &     46.00 &     -46.00 &    1.00 &              6.00 &          1.00 \\
25\%   & 14,836.39 &  19,274.31 &    3.00 &             12.00 &          2.00 \\
50\%   & 21,592.43 & 115,241.45 &   12.00 &             18.00 &          5.00 \\
75\%   & 25,054.40 & 124,243.35 &   16.00 &             20.00 &          8.00 \\
max   & 55,953.37 & 777,548.67 &   18.00 &             42.00 &         14.00 \\
\bottomrule
  \end{tabular}
  \end{adjustbox}
  \caption{Distributions for suppression attacker clusters.}
  \label{tbl:suppression-cluster-distributions}
\end{table}

\noindent
\newline
\textbf{Competition.}
We found that suppression attacks only targeted 15 unique contracts, which are listed in Table~\ref{tbl:suppressed_contracts}.
We can see that only the contracts Last Winner, FoMo3Dlong, and Peach Will were targeted by different attacker clusters.
We searched through all the attacks for blocks where any of these three contracts were the victims and more than one attacker cluster was targeting the same victim.
We found only one case where bot contract 0xDd9fd6b6F8f7ea932997992bbE67EabB3e316f3C started an attack interrupting another attack from bot contract 0xd037763925C23f5Ba592A8b2F4910D051a57A9e3 targeting Last Winner on block 6,232,122. 

\subsection{Summary}

\noindent
In the following, we summarize our previous findings and compare the different types of frontrunning attacks in terms of structure, competition, cost, profit, bot triggers, bot activity, and trends.

\noindent
\newline
\textbf{Structure.}
As shown in Figure~\ref{fig:frontrunning}, the difference between each attack type is the number of transactions the attacker employs and where the attacker places them in a block relative to the victim. For displacement, the attacker needs to place only one transaction before the victim. For insertion, the attacker creates a sandwich of two transactions around the victim's transaction. Finally, for suppression, the attacker must delay the victim's transaction with one or more transactions.

\noindent
\newline
\textbf{Competition.} Attackers can interrupt each other depending on the structure of the attack. For displacement, the attacker only sends one transaction before the victim, so the only way for another attacker to interrupt the attack is to place another transaction before the attacker (i.e., with a higher gas price than the victim and the attacker). Moreover, note that the second attacker could be aiming at the same victim or could be considering the first attacker as the victim. In insertion, competition is more complex since one or more transactions can interfere between the three transactions of a sandwich. Additionally, one attacking transaction can take the role of the victim in another sandwich (i.e., when the attacking transaction moves so many funds that it is considered a whale transaction for other attackers). Finally, the suppression case is even easier to interrupt given the number of transactions it involves over an extended range of blocks. Interestingly, the results from our heuristic show that interruptions from regular lottery participants caused most of the failed attacks.

\begin{figure*}
\begin{center}
\begin{minipage}{.5\textwidth}
    \centering
    \includegraphics[width=\linewidth]{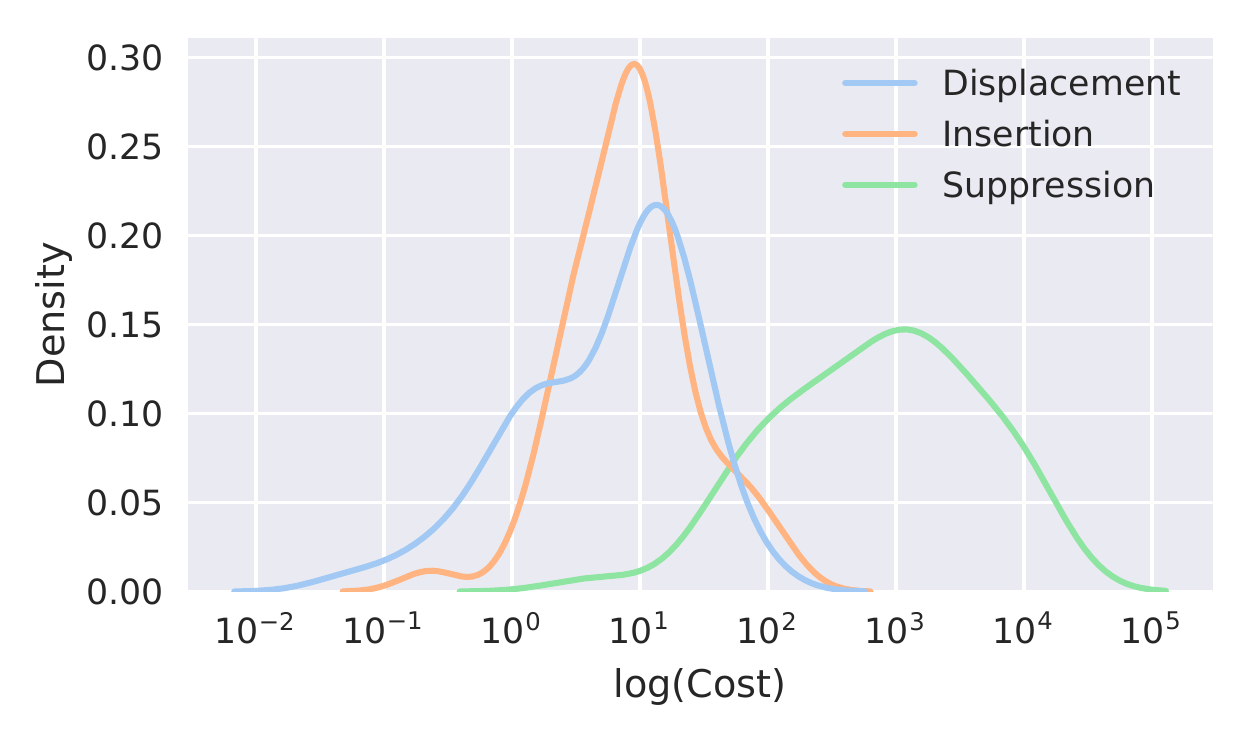}
\end{minipage}%
\begin{minipage}{.5\textwidth}
    \centering
    \includegraphics[width=\linewidth]{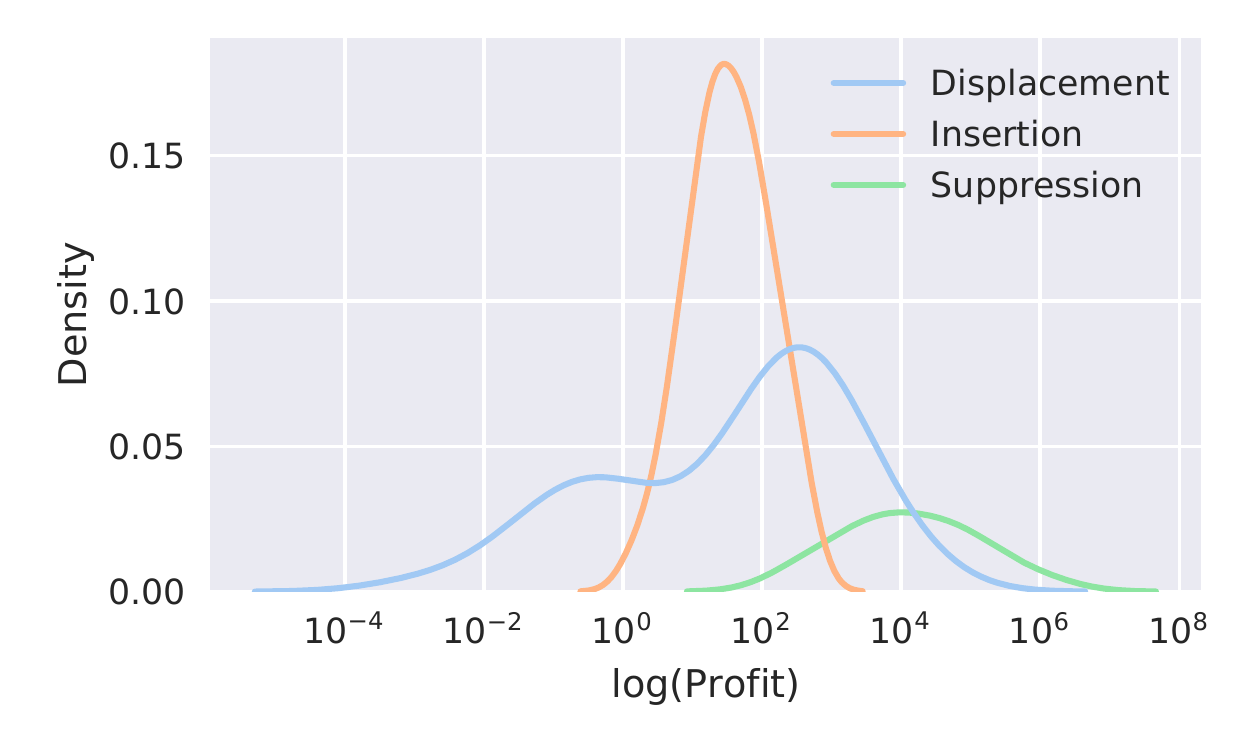}
\end{minipage}
\end{center}
\caption{\label{fig:distribution-comparison} Cost (left) and profit (right) distributions in logarithmic scale.}
\end{figure*}

\noindent
\newline
\textbf{Cost.} We present the distribution of attack cost for each attack type in Tables~\ref{tbl:displacement-attack-distributions}, ~\ref{tbl:insertion-attack-distributions} and~\ref{tbl:suppression-attack-distributions}. In Figure~\ref{fig:distribution-comparison} (left), we present the three cost distributions all together. We employed a logarithmic scale on the x-axis because the high density of displacement cost around zero as well as the large dispersion of suppression costs, makes the visualization hard to interpret. However, using this logarithmic scale, we cannot visualize the actual cost, but we can see that suppression attacks tend to be more expensive and have more diverse costs.

\noindent
\newline
\textbf{Profit.} Similar to the cost, we present the distribution of attack profit for each attack type in Tables~\ref{tbl:displacement-attack-distributions}, ~\ref{tbl:insertion-attack-distributions} and~\ref{tbl:suppression-attack-distributions}. In Figure~\ref{fig:distribution-comparison} (right), we present the three profit distributions all together. Similar to the cost, we employed a logarithmic scale on the x-axis because the high density of insertion profit around zero as well as the large dispersion of suppression profits, makes the visualization hard to interpret. In this scale, we cannot visualize the actual profit, but we can see that displacement attacks tend to be more profitable than insertion attacks and that suppression attacks tend to be more profitable than the other two. Additionally, the displacement profit distribution seems to have two modes.

\noindent
\newline
\textbf{Bot Triggers.} Bots are triggered by transactions that appear in the pool of pending transactions, which on the other hand reflects user activity. Thus, bots respond to actions performed by human users. For instance, in the case of displacement these triggers can be users accessing smart contracts that do not have proper access control. For insertion, bots are typically triggered by large trades that users commit on decentralized exchanges. Finally, for suppression, bots are triggered when the prize pool of a lottery or gambling contract has reached a significant amount of value, which makes running a suppression attack lucrative.

\begin{figure}[H]
\centering
\includegraphics[width=0.95\linewidth]{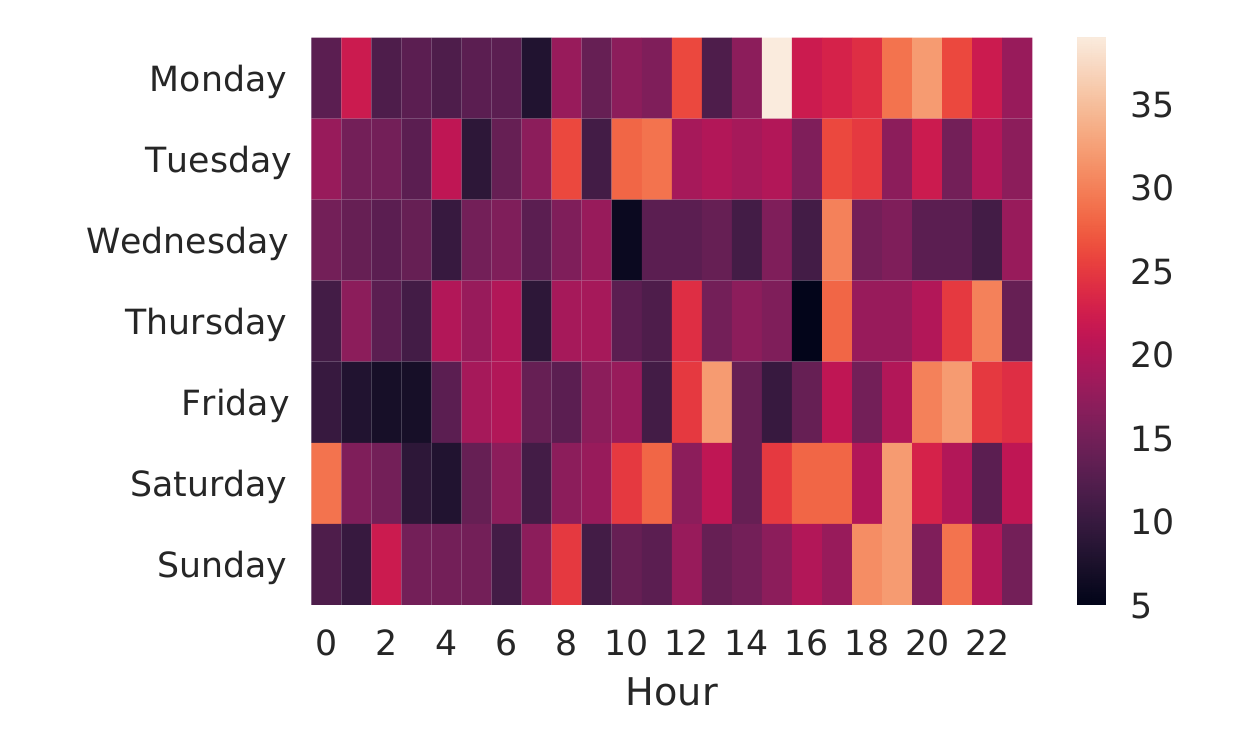}
\includegraphics[width=0.95\linewidth]{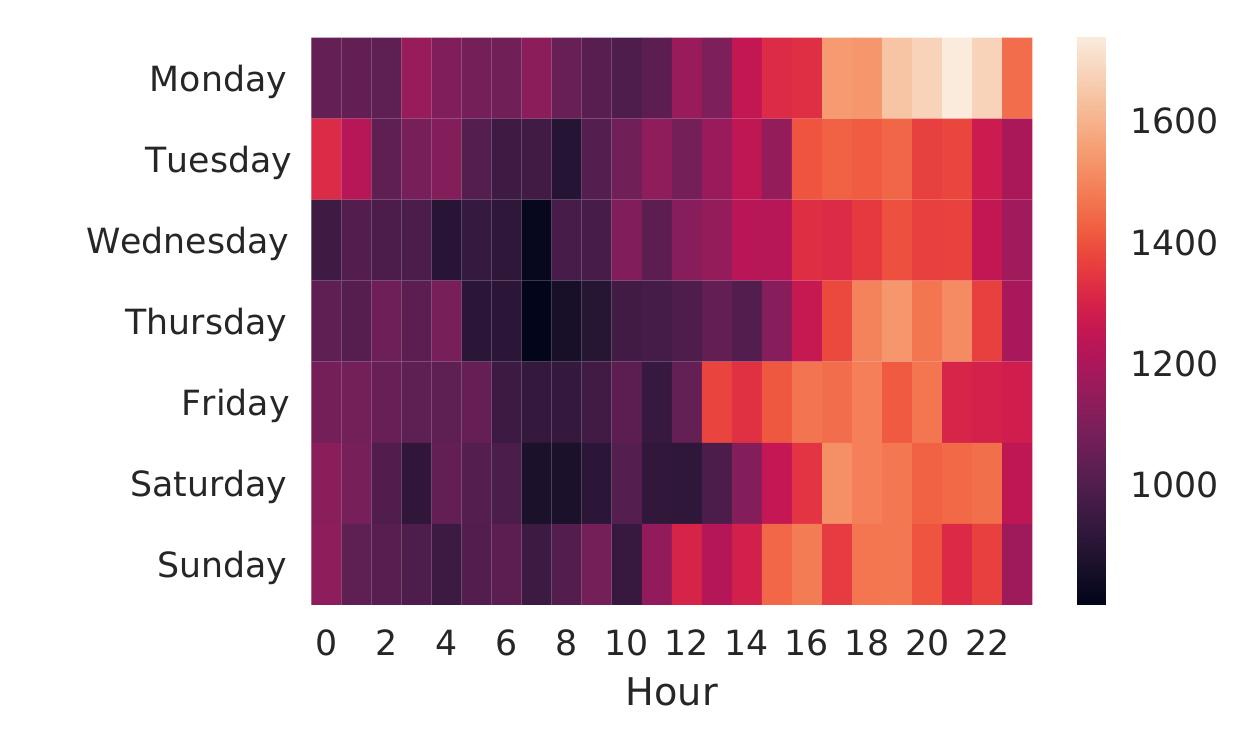}
\includegraphics[width=0.95\linewidth]{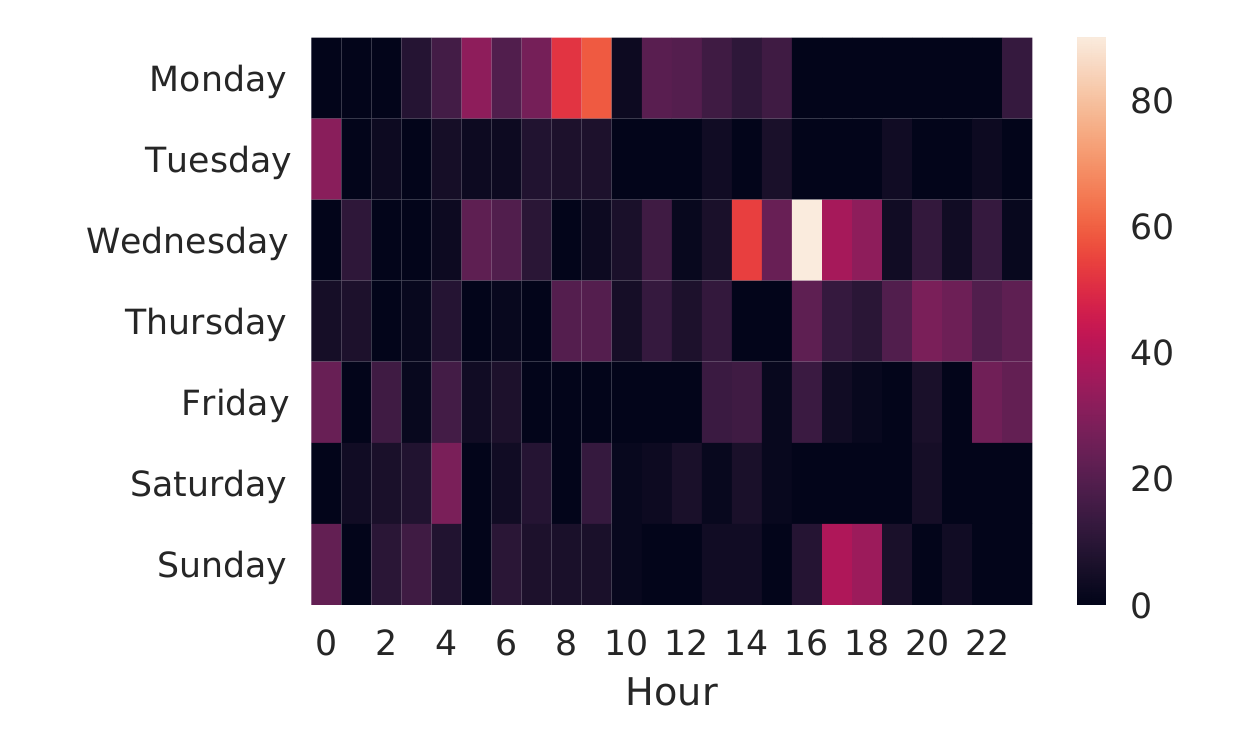}
\caption{Number of attacks by weekday and hour for displacement (top), insertion (middle), and suppression (bottom), following the UTC timezone.}
\label{fig:weekday-hour}
\end{figure}

\noindent
\newline
\textbf{Bot Activity.} Figure~\ref{fig:weekday-hour} describes the number of attacks by weekday and hour for displacement, insertion, and suppression, respectively, using Coordinated Universal Time (UTC) as timezone. We can see that the distribution for displacement appears to be random. 
For insertion, our results indicate higher bot activity overlapping with evening hours in the northern hemisphere, with highest activity between five and midnight. One plausible explanation is that transactions vulnerable to insertion attacks correspond to human-initiated trading on the blockchain.
The evening activity can be explained by the fact that most people have more time to do trading on decentralized exchanges at the end of the day (e.g. after work or after dinner).
However, as discussed previously, user activity triggers bots, and users belong to different parts of the world with different timezones, so it is hard to draw any conclusions. We leave it to future work to validate whether the increase of trading activity on decentralized exchanges has led to the increase of insertion frontrunning attacks and whether most users engage in trading activities at the end of the day. Finally, there is a slightly higher activity on Wednesdays for suppression, but we are unsure if the reason depends on a particular lottery (e.g. advertisement) or if this is just a coincidence due to our small sample size of detected suppression attacks.

\begin{figure}
\centering
\includegraphics[width=0.95\linewidth]{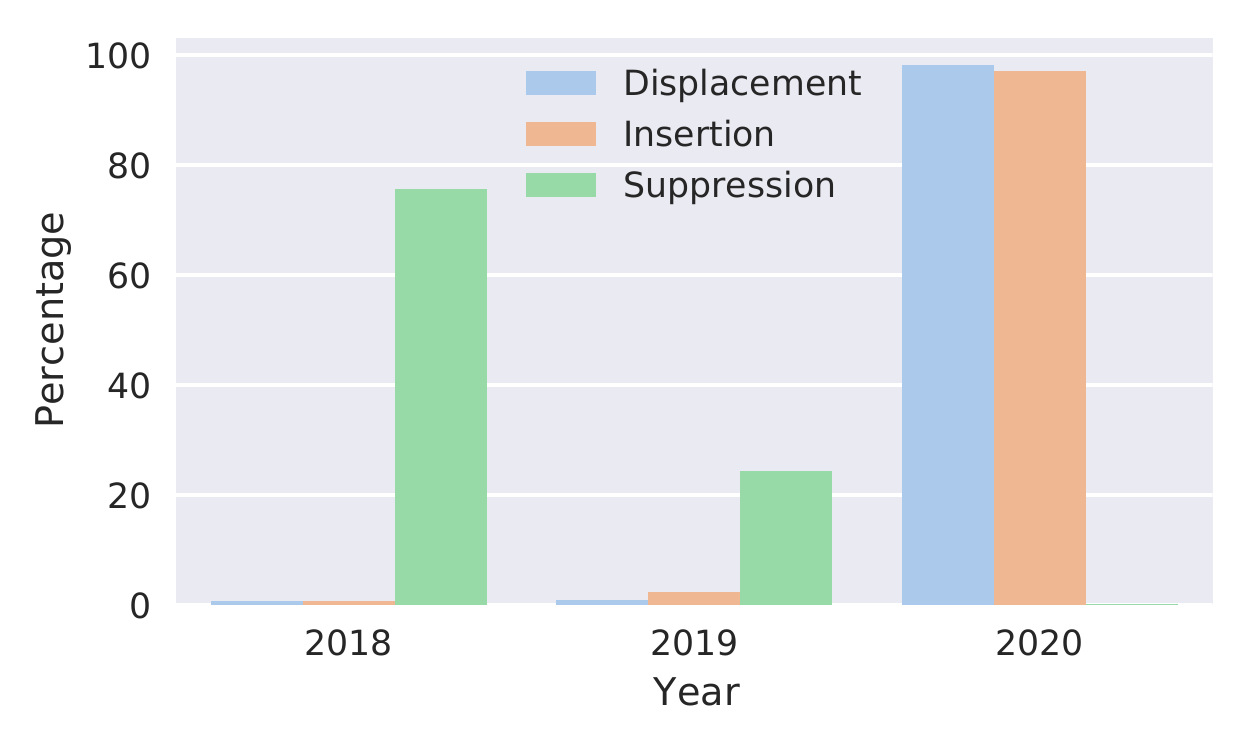}
\caption{Percentage of attacks by year.}
\label{fig:year-percentage}
\end{figure}

\noindent
\newline
\textbf{Trends.} The number of attacks has a very different magnitude for each attack type: ~2K for displacement, ~197K for insertion and only 50 for suppression. This difference makes it hard to visualize how the amount of attacks changes overtime for all the attacks at the same time. For that reason, in Figure~\ref{fig:year-percentage}, we present the percentage of attacks by year for each type of attack. We cannot compare the absolute values in the y-axis, but we can see how suppression attacks decreased over the years, and how both, displacement and insertion, mostly appear in 2020.

\section{Discussion}

In this section, we discuss the implications of frontrunning and why existing mitigation techniques are not effective.

\subsection{Implications of Frontrunning}

Daian \etal \cite{daian2019flash} emphasize that miners could engage in frontrunning activities to maximize or increase their profits. This will most likely be the case when EIP-2878 becomes accepted and the current static block award drops from 2 ETH to 0.5 ETH \cite{eip2878}. However, at the moment miners are already profiting indirectly from frontrunning activities performed by non-miners, since the high gas prices that those non-miners pay end up being for the miners in the form of transaction fees. Thus, miners are incentivized to allow frontrunning. Our results show that miners already earned more than 300K USD from transaction fees payed by the attackers performing insertion frontrunning attacks.
While transaction fees in January 2018 only represented 9\% of the monthly revenue of a miner, in January 2021 nearly 40\% of the monthly revenue came from transaction fees \cite{miner_profit}.
Thus, besides attackers, we also concluded that miners profit from frontunning attacks. However, attackers and miners are not the only entities that profit from frontrunning. Take the example of Uniswap. In general, Uniswap takes a 0.3\% fee on every transaction. This fee is divided between the liquidity providers, proportionally to their share. For example, if you provide 50\% of the liquidity, then you will earn 50\% of the collected fee.
Thus, liquidity providers profit from every trade performed by frontrunners.
However, frontrunning attacks can also have some severe implications for normal users in general. For instance, due to multiple attackers trying to frontrun other attackers via gas price auctions, they temporarily push the average gas prices of the network and force users that do not engage in frontrunning to either pay higher transaction fees or wait longer for their transactions to be mined. 
This becomes a vicious circle where once again the miners profit from the fact that benign users have to pay higher transaction fees for their transactions to be mined. Thus, the more attackers engage in frontrunning, the more it will have an impact on benign users.
Another issue is suppression, which prevents blocks to be used or filled in an optimal way. Ethereum already struggles with a low transaction throughput \cite{throughput} and suppression attacks only amplify the issue. Suppression attacks can cause the network to congest and decentralized applications to stop working properly.

\subsection{Limitations of Existing Mitigations}

There are currently two main reasons why frontrunning is conceivable on public blockchains such as Ethereum.
The first reason is the lack of transaction confidentiality. Every node in the network, not just miners, can observe all the transactions in the clear before they are mined. The fact that transactions are transparent to everyone is undoubted one of the major advantages of a public blockchain, however the content and purpose of a transaction should only be visible to everyone once it has been mined.
The second reason is the miner’s ability to arbitrarily order transactions. This puts a lot of power into the hands of miners. Miners can decide to censor transactions or change the order of transactions such that they make the most profit. The idea to order transactions based on the gas price sounds rational at first, however this also introduces determinism in a way that can be manipulated by outsiders.
A suitable mitigation technique must address these two issues, but it must also be efficient in terms of costs for the users, provide fair incentives for miners to continue mining transactions, and be adoptable by everyone and not just by a special group of participants.
In our study, we observed that most frontrunning is happening on DEXes, since the risk of failure is low compared to the amount of profit that can be made.
Uniswap, the DEX most affected by frontrunning, is aware of the frontrunning issue and proposes a slippage tolerance parameter that defines how distant the price of a trade can be before and after execution.
The higher the tolerance, the more likely the transaction will go through, but also the easier it will be for an attacker to frontrun the transaction. 
The lower the tolerance, the more likely the transaction will not go through, but also the more difficult it will be for an attacker to frontrun the transaction.
As a result, Uniswap's users find themselves in a dilemma. 
Uniswap suggests by default a slippage tolerance of 0.5\% in order to minimize the likelihood that users become victims of frontrunning.
However, in this work we prove that the slippage tolerance does not work as we measured over 180K attacks against Uniswap.
Hence, other mitigations to counter frontrunning are needed.
Bentov \etal \cite{bentov2019tesseract} present \textsc{Tesseract}, an exchange that is resistant to frontrunning by leveraging a trusted execution environment. However, their design follows a centralized approach and requires users to have hardware support for trusted execution.
Breidenbach \etal \cite{breidenbach2018enter} proposed LibSubmarine\cite{libsubmarine}, an enhanced commit-and-reveal scheme to fight frontrunning. However, in the case of Uniswap, LibSubmarine would require three transactions to perform a single trade, making it cumbersome and relatively expensive for users to trade.

\section{Related Work}

Daian \etal researched frontrunning attacks from an economical point of view by studying gas price auctions \cite{daian2019flash}. Moreover, by modeling actions of bots using game theory, and framing the problems in terms of a Nash equilibrium for two competing agents, the authors demonstrated that DEXes are severely impacted by two main factors: the high latency required to validate transactions, which opens the door to timing attacks, and secondly the miner driven transaction prioritization based on miner extractable value. The mix of these two factors leads to new security threats to the consensus-layer itself, independent of already existing ones \cite{bonneau2016buy,eyal2014majority}.
However, the authors only focused on detecting frontrunning on DEXes and in real time, without scanning the entire blockchain history for evidence of frontrunning. 
Our work builds on the taxonomy defined by Eskandari \etal \cite{eskandari2019sok}, which introduces three different types of frontrunning: displacement, insertion, and suppression. Despite illustrating a few concrete examples and discussing several mitigation techniques, the authors did not analyze the prevalence of frontrunning attacks in the wild.
Zhou \etal \cite{zhou2020high} estimated the potential effect of frontrunning on DEXes but limited their analysis only to insertion attacks on a single exchange. Their study estimated the theoretical profit that could have been made if users would have engaged in fruntrunning attacks, but did not back their conclusion with real observed data. Compared to their work, we perform real world measurements not only for insertion attacks, but for the complete spectre of attack types (\ie displacement, insertion, and suppression).
Besides studying frontrunning, a few mitigation techniques have also been proposed to counter frontrunning.
For instance, Kelkar \etal proposed a consensus protocol to achieve transaction order-fairness \cite{kelkar2020order}.
Breidenbach \etal \cite{breidenbach2018enter} proposed LibSubmarine\cite{libsubmarine}, an advanced commit-and-reveal scheme to fight frontrunning at the application layer. 
Bentov \etal \cite{bentov2019tesseract} present \textsc{Tesseract}, an exchange that is resistant to frontrunning by leveraging a trusted execution environment.
Finally, Kokoris \etal \cite{kokoriscalypso} describe \textsc{Calypso}, a blockchain that is resistant to frontrunning due to private transactions.
Unfortunately, none of these techniques are broadly adopted as they are either not compatible with the Ethereum blockchain or because they are too costly.
Another important side-effect of decentralized finance is the emergence of flash loans \cite{werner2021sok}. 
Wang \etal \cite{wang2020towards} discuss a methodology to detect flash loans using specific patterns and heuristics.
We leave it to future work to study the implications of flash loans in the context of frontrunning.

\section{Conclusion}

In this work, we investigated the prevalence of frontrunning attacks in Ethereum.
To the best of our knowledge, we are the first to present a methodology to efficiently measure the three different types of frontrunning attacks: \emph{displacement}, \emph{insertion}, and \emph{suppression}.
We performed a large-scale analysis on the Ethereum blockchain and identified 199,725 attacks with an accumulated profit of over 18.41M USD for the attackers. 
We also discussed implications of frontrunning and found that miners profit from frontrunning practices. We found that miners already made a profit of more than 300K USD from transaction fees payed by frontrunners.
We hope that we shed with this work some light on the predators of Ethereum's dark forest by providing evidence that frontrunning is both, lucrative and a prevalent issue.

\section*{Acknowledgments}

We would like to thank the anonymous reviewers and Shaanan Cohney for their valuable comments and feedback.
We also thankfully acknowledge the support from the RIPPLE University Blockchain Research Initiative (UBRI) and the Luxembourg National Research Fund (FNR) under grant 13192291.

\bibliographystyle{plain}
\bibliography{references}

\end{document}